\documentclass[a4paper]{article}
\usepackage{fullpage}

\usepackage{authblk}
\usepackage{amsmath}
\usepackage{amsfonts}
\usepackage{mathtools}
\usepackage{amssymb}
\usepackage{amsthm}
\usepackage{mhchem}
\usepackage{url}
\usepackage{siunitx}
\usepackage{rotating}
\usepackage{makecell}
\usepackage{hhline}
\usepackage{caption}
\usepackage{subcaption}
\usepackage[dvipsnames]{xcolor}
\usepackage{bbm}

\usepackage{natbib}
\graphicspath {{figures/}}

\newtheorem{thm}{Theorem}
\newtheorem{defi}[thm]{Definition}
\newtheorem{ass}{Assumption}
\newtheorem{prop}[thm]{Proposition}
\newtheorem{ex}[thm]{Example}

\newtheorem{cor}[thm]{Corollary}
\newtheorem{rem}[thm]{Remark}

\renewcommand\epsilon{\varepsilon}
\renewcommand\subset{\subseteq}

\newcommand\N{\mathbb{N}}

\newcommand\R{\mathbb{R}}

\newcommand\E{\mathbb{E}}
\newcommand\fave{f_{\text{AVE}(X \to Y)}}
\newcommand\favehat{\hat{f}^{nm}_{\text{AVE}(X \to Y)}}

\newcommand{\B}[1]{\mathbf{#1}}

\newcommand{\given}{\, \vert \,}
\newcommand{\st}{\, : \,}

\DeclarePairedDelimiterX{\card}[1]{\lvert}{\rvert}{#1}
\DeclarePairedDelimiterX{\norm}[1]{\lVert}{\rVert}{#1}

\DeclareSymbolFont{largesymbolsA}{U}{txexa}{m}{n}
\DeclareMathSymbol{\varprod}{\mathop}{largesymbolsA}{16} %

\newcommand\indep{\protect\mathpalette{\protect\independenT}{\perp}}
\def\independenT#1#2{\mathrel{\rlap{$#1#2$}\mkern2mu{#1#2}}}

\usepackage[framemethod=TikZ]{mdframed} 
\usepackage{tikz}
\usetikzlibrary{arrows}
\usetikzlibrary{positioning}
\newdimen\arrowsize
\pgfarrowsdeclare{arcsq}{arcsq}
{
  \arrowsize=0.2pt
  \advance\arrowsize by .5\pgflinewidth
  \pgfarrowsleftextend{-4\arrowsize-.5\pgflinewidth}
  \pgfarrowsrightextend{.5\pgflinewidth}
}
{
  \arrowsize=1.5pt
  \advance\arrowsize by .5\pgflinewidth
  \pgfsetdash{}{0pt} %
  \pgfsetroundjoin   %
  \pgfsetroundcap    %
  \pgfpathmoveto{\pgfpoint{0\arrowsize}{0\arrowsize}}
  \pgfpatharc{-90}{-140}{4\arrowsize}
  \pgfusepathqstroke
  \pgfpathmoveto{\pgfpointorigin}
  \pgfpatharc{90}{140}{4\arrowsize}
  \pgfusepathqstroke
}

\tikzset{every picture/.style={line width=0.6pt}}
\tikzset{every picture/.style={outer sep=.4mm}}
\tikzstyle{graphnode} = 
   [circle,draw=black,minimum size=25pt,text centered,inner sep=0.2mm] 
\tikzstyle{observed}   =[graphnode,fill=white,text=black]
\tikzstyle{unobserved}   =[graphnode,fill=white,text=black,style=dashed]
\tikzstyle{graphnodesmall} = 
   [circle,draw=black,minimum size=20pt,text centered,inner sep=0.2mm] 
\tikzstyle{observedsmall}   =[graphnodesmall,fill=white,text=black]
\tikzstyle{unobservedsmall}   =[graphnodesmall,fill=white,text=black,style=dashed]
\tikzstyle{graphnodestiny} = 
   [circle,draw=black,minimum size=15pt,text centered,inner sep=0.2mm] 
\tikzstyle{observedtiny}   =[graphnodestiny,fill=white,text=black]
\tikzstyle{unobservedtiny}   =[graphnodestiny,fill=white,text=black,style=dashed]

\definecolor{light-gray}{gray}{0.97}
\definecolor{commentgray}{gray}{0.5}
 
\DeclareRobustCommand\sampleline[1]{%
  \tikz\draw[#1] (0,0) (0,\the\dimexpr\fontdimen22\textfont2\relax)
  -- (1em,\the\dimexpr\fontdimen22\textfont2\relax);%
}

\title{
\bf
Towards Causal Inference for Spatio-Temporal Data: \\
Conflict and Forest Loss in Colombia
}
\author{Rune Christiansen$^\dagger$\thanks{corresponding author, e-mail: krunechristiansen@gmail.com} \quad Matthias Baumann$^\ddagger$ \quad Tobias Kuemmerle$^{\ddagger \diamond}$ \\ Miguel D. Mahecha$^{\S \P}$ \quad Jonas Peters$^\dagger$}
\affil{
\normalsize{
\textit{$^\dagger$Department of Mathematical Sciences, University of Copenhagen} \\ 
\textit{$^\ddagger$Geography Department, Humboldt-Universität zu Berlin} \\ 
\textit{$^\diamond$Integrative Research Institute on Transformations of Human-Environment Systems (IRI THESys), Humboldt-Universität zu Berlin} \\
\textit{$^\S$Remote Sensing Center for Earth System Research, Leipzig University} \\ 
\textit{$^\P$German Centre for Integrative Biodiversity Research (iDiv)}
}
}
\date{\today}

\renewcommand\P{\mathbb{P}}

\begin{document}
\maketitle

\begin{abstract}%
In many data scientific problems, we are interested in inferring 
causal relationships 
in the data generating mechanism. Here, 
we consider the following real-world question: how has the 
Colombian conflict influenced tropical forest loss? There is evidence 
for both enhancing and reducing impacts. Answering 
such questions requires the use of causal models. In this work, we propose a 
class of causal models for spatio-temporal stochastic processes. It 
allows us to formally define and quantify the causal effect of a vector 
of covariates $X$ on a real-valued response $Y$, even if the causal 
background knowledge is incomplete. We introduce a procedure for 
estimating causal effects, and a non-parametric hypothesis test for 
these effects being zero. The proposed methods do not make strong 
distributional assumptions, and allow for arbitrarily many latent 
confounders, given that these confounders do not vary across time (or, 
alternatively, they do not vary across space). When applying our causal 
methodology to the problem of conflict and forest loss, using
data from 2000 to 2018, we find a reducing but 
insignificant causal effect of conflict on forest loss. Regionally, both 
enhancing and reducing effects can be identified.
Our theoretical findings are supported by simulations, and code is available online.
\end{abstract}

\section{Introduction}
\subsection{Spatio-temporal data analysis}
In principle, all data are spatio-temporal data: 
Any observation of any phenomenon 
occurs at a particular point in space and time. If information on the 
spatio-temporal origin of data is available, 
this information
can be exploited 
for statistical modeling 
in various ways; this is the study of 
spatio-temporal statistics 
\citep[e.g.,][]{cressie2015statistics,wikle2019spatio}. 
Spatio-temporal statistical models find their application in many 
environmental and sustainability sciences, 
and have been used, for example, 
for the analysis of biological growth patterns \citep{chaplain1999growth}, 
to model meteorological fields \citep{bertolacci2019climate},
or to assess the development of land-use change \citep{liu2017spatio} 
and sea level rise \citep{zammit2015multivariate}. 
They are 
used in epidemiology 
for prevalence mapping of infectious diseases \citep{giorgi2018geostatistical}, 
and play a key role in socio-economic research,
for example, 
for the modeling 
of housing prices \citep{holly2010spatio}, 
or for election forecasting \citep{pavia2008election}. 
In almost all of these domains, 
the abundance of 
spatio-temporal
data 
has increased rapidly 
over the last decades.
Several 
advances 
aim to 
improve the accessibility 
of such datasets, 
e.g., via `data cube' approaches 
\citep[e.g.,][]{nativi2017view, mahecha2020earth}.

Most spatio-temporal statistical models are models for the observational distribution, 
that is, 
they model processes that are
passively observed. 
By allowing for spatio-temporal trends and dependence structures, 
such models can be accurate descriptions of complex processes, 
and have proven to be effective tools for 
spatio-temporal prediction,
inference and forecasting \citep[e.g.,][]{wikle2019spatio}. 
However, to answer interventional questions such as 
``How does a certain policy change affect land-use patterns?'', 
we require a model for the intervention distribution, that is, 
for data generated under a change in the data generating mechanism --- we 
require a \textit{causal} model for the data generating process.

\subsection{Causality}
Causal models can be used to quantify causal relations between random variables, 
(e.g., by analyzing the change in expected value of $Y$ when intervening on $X$). 
However, existing causal models do not apply well to our setting, since they are either 
restricted to independent and identically distributed (i.i.d.) or time series data,
or they rely on various assumptions which cannot be expected to hold in our application (see Appendix~\ref{app:existing_causal_methods}
for a review of existing work).
In this paper, we introduce a novel class of causal models for 
multivariate spatio-temporal stochastic processes. 
A spatio-temporal dataset may then be viewed as 
a single realization from such a model, 
observed at discrete points in space and time.
The full causal structure among all variables of a spatio-temporal process 
can hardly be fully specified. 
In practice, however, a full causal specification may also not be necessary:
we are often interested in quantifying only certain causal relationships, 
while being indifferent 
to other parts of the causal structure. 
The causal models introduced in this paper
are well adapted to such settings. 
They allow us to model
a causal influence of a vector of covariates $X$
on a target variable $Y$ while leaving other 
parts of the causal structure unspecified.
In particular, the models accommodate largely unspecified autocorrelation patterns in the response variable, 
which are a common phenomena in spatio-temporal data.

The introduced framework allows us to formally talk
about causality in a spatio-temporal context
and can be used to construct well-defined 
targets of inference.
As an example, we define 
the intervention effect 
(`causal effect') of 
$X$ on 
$Y$.
We show 
that
this
effect can be estimated 
from observational spatio-temporal data
and introduce a corresponding estimator.
We prove consistency and verify this finding by a simulation experiment.
We furthermore construct a non-parametric 
hypothesis test for the causal effect being zero. Our methods 
do not rely on any distributional assumptions on the data generating process.
They furthermore
allow for the 
influence of arbitrarily many latent confounders if these confounders do not vary across 
time. 
In principle, our method also 
allows to analyze 
problems where temporal and spatial 
dimensions are interchanged, meaning that confounders may vary in time but remain 
static across space.

Our work has been motivated by the following application.

\subsection{Conflict and forest loss in Colombia} \label{sec:conflict}
Tropical forests are rich in biodiversity \citep{kreft2007global},
store large amounts of carbon \citep{avitabile2016integrated}, 
play an important role 
in climate-regulation, and provide livelihoods to millions of people \citep{lambin2011global}. 
Yet, tropical forests
continue to be under pressure
due to agricultural expansion 
\citep{
angelsen1999rethinking}, 
mining \citep{sonter2017mining}, timber harvest \citep{pearson2014carbon} 
or urban expansion \citep{defries2010deforestation}. 
A problem that is still only partly understood is the interaction between forest loss 
and armed conflicts \citep{baumann2016impacts}, %
which are frequent events 
in tropical areas \citep{pettersson2015armed}. 
Armed conflict may have both positive 
and negative impacts on forest loss. 
On one hand, 
conflict can lead to increasing pressure on forests, as 
timber 
may finance warfare activities 
\citep{harrison2015blood}. Also, reduced law enforcement in conflict regions may
lead to 	plundering of natural resources 
leading to increasing forest loss \citep{butsic2015conservation}. 
On the other hand, the outbreak of armed conflicts can reduce pressure 
on forest resources, e.g., %
when economic and political insecurity
interrupt large-scale mining activities, or when economic sanctions stop international
timber trade \citep{lebillon2000political}. Investors may furthermore be hesitant to invest 
in agricultural activities \citep{collier2000economic}, thereby reducing the pressure on 
forest areas compared to peace times \citep{gorsevski2012analysis}. 

Here, we focus on the specific case of Colombia, where an armed conflict 
has been present for over 50 years, causing more than 200,000 fatalities, until a 
peace agreement was reached in 2016. %
There is evidence that increased forest loss can be, at least regionally, attributed to the armed conflict \citep{castro2017land, landholm2019diverging}. 
At the same time, there are also arguments suggesting that the pressure on forests was partially
reduced when armed conflict prevented logging \citep{davalos2016deforestation}.
Most papers report evidence that both positive and negative impacts of conflict 
on forest loss 
may happen in parallel, depending on the local conditions
\citep[e.g.,][]{sanchez2013consequences, castro2017land}. 
We believe that a 
purely data-driven
approach can be a useful addition 
to this debate.

In our analysis, we use a 
dataset containing 
the following variables.
\begin{itemize}
\setlength\itemsep{0em}
\item $X_{s}^t:$ binary conflict indicator for location $s$ at year $t$.
\item $Y_{s}^t:$ absolute forest loss in location $s$ from year $t-1$ to year $t$, measured in $\si{\km}^2$.
\item $W_{s}^t:$ distance from location $s$ to the closest road, measured in $\si{\km}$.
\end{itemize}
Data are annually aggregated, covering the years from 2000 to 2018, and spatially explicit 
at a $10 \si{\km} \times 10 \si{\km}$-resolution. We provide a detailed description of 
the data processing in Section~\ref{sec:conflict_cont}. 
\begin{figure}[t]
\centering
\includegraphics[width = .7\linewidth]{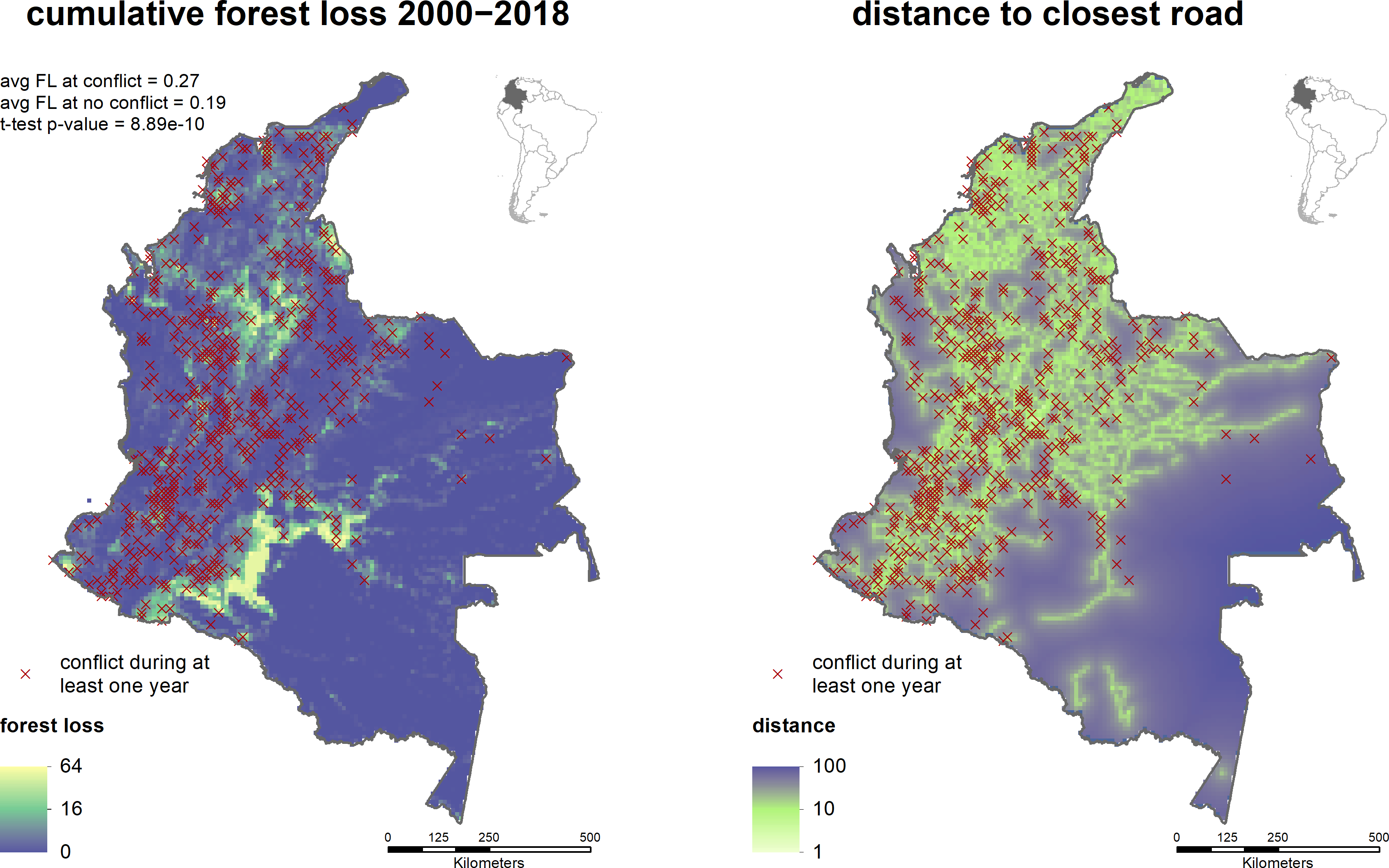}
\caption{
Temporally aggregated summary of the dataset described in Section~\ref{sec:conflict}. 
Conflicts are predictive of exceedances in forest loss (left), but this dependence is partly induced by a common 
dependence on accessibility, which we measure by the mean distance to a road (right). 
Failing to account for this variable 
and other confounders
biases 
our estimate of the causal influence of conflict on forest loss. 
}
\label{fig:col_fl}
\end{figure}
A summary of the dataset can be seen in Figure~\ref{fig:col_fl}.
Visually, there is a strong positive dependence between the occurrence of a conflict %
and the loss of forest canopy. This observation is supported by simple 
summary statistics: the average forest loss across measurements classified as conflict events
significantly exceeds that from non-conflict events by almost 50\% 
(Figure~\ref{fig:col_fl}, left), conforming previous findings
\citep[e.g.,][]{landholm2019diverging}.
When seeking a causal explanation for the observed data, however,
we regard 
such an analysis as flawed in two ways.
First, both conflicts and forest loss predominantly occur in areas with high accessibility (Figure~\ref{fig:col_fl}, right), 
indicating that the potential causal effect of $X$ on $Y$ is confounded by $W$.
In fact, we expect the existence of several other confounders (e.g., population density, market infrastructure, etc.), many of which may be unobserved. 
Failing to account for confounding variables %
leads to biased estimates of the causal effect. 
Second, strong spatial dependencies in $X$ and $Y$ reduce the effective sample size, 
and a standard t-test thus exaggerates the significance of the observed difference in sample averages. 
To test hypotheses about $X$ and $Y$, 
we need statistical tests which are adapted to the spatio-temporal nature of data.

\subsection{Contributions and structure of the paper} \label{sec:contri}
Apart from the case study, this paper
contains 
three main theoretical contributions: 
the definition of a causal model
for spatio-temporal data, a method 
for estimating causal effects (both Section~\ref{sec:quant}), 
and a hypothesis test %
for the overall existence of such effects (Section~\ref{sec:testing}).
Section~\ref{sec:conflict_cont} applies
our methodology to the above example.
All data used for our analysis are publicly available. A description of how it can be obtained, 
along with an implementation of our method and reproducing scripts for all our figures and results, 
can be found at \url{github.com/runesen/spatio_temporal_causality}. All proofs are contained in Appendix~\ref{app:proofs}.

\section{Quantifying causal effects for spatio-temporal data} \label{sec:quant}
A spatio-temporal dataset may be viewed as a single 
realization of a spatio-temporal 
stochastic process, 
observed at discrete points in space-time.
In this section, we provide a formal framework to quantify causal 
relations among the components of a multivariate spatio-temporal process,
and show how to estimate causal effects from observational 
data.

\subsection{Causal models for spatio-temporal processes} \label{sec:causal_models}
Throughout this section, let $(\Omega, \mathcal{A}, P)$ be some background probability space. 
A $p$-dimensional \emph{spatio-temporal process} $\B{Z}$ is a random variable taking values in the 
sample space $\mathcal{Z}_p$ of all $(\mathcal{B}(\R^2 \times \N), \mathcal{B}(\R^p))$-measurable functions, 
where $\mathcal{B}(\cdot)$ denotes the Borel $\sigma$-algebra. 
We equip $\mathcal{Z}_p$ with the $\sigma$-algebra $\mathcal{F}_p$, defined as the smallest $\sigma$-algebra 
such that 
for all $B \in \mathcal{B}(\R^p)$,
the mapping $\mathcal{Z}_p \ni z \mapsto \int_B z(x) dx$ is
$(\mathcal{F}_p, \mathcal{B}(\R))$-measurable. 
The induced probability measure $\P$ on the measurable space 
$(\mathcal{Z}_p, \mathcal{F}_p)$, for every $F \in \mathcal{F}_p$ defined 
by $\P(F) := P(\B{Z}^{-1}(F))$, is said to be the \emph{distribution of $\B{Z}$}. 
Throughout this paper, we 
use the notation $Z_s^t$ to denote the random vector obtained from marginalizing $\B{Z}$ at location $s$ and time point $t$. 
We use $\B{Z}_s$ for the time series 
$(Z_{s}^t)_{t \in \N}$, $\B{Z}^t$ for the spatial process $(Z_s^t)_{s \in \R^2}$, and 
$\B{Z}^{(S)}$ for the spatio-temporal process corresponding to the coordinates in $S \subset \{1, \dots, p\}$. 
We call a spatio-temporal process \emph{weakly stationary} if the marginal distribution 
of $Z_s^t$ is the same for all $(s,t) \in \R^2 \times \N$, and \textit{time-invariant} 
if $\P(\B{Z}^1 = \B{Z}^2 = \cdots ) = 1$. 

Multivariate spatio-temporal processes are used for the joint modeling of
different phenomena, each of which corresponds 
to a coordinate process. 
Let us consider a decomposition of these coordinate processes into disjoint `bundles'.
We are interested in specifying causal relations 
among these bundles
while leaving the causal structure 
among variables within each bundle unspecified. 
Similarly to a graphical model \citep{lauritzen1996graphical}, our approach relies on a factorization of the joint 
distribution of $\B{Z}$ into a number of components, each of which models the conditional 
distribution for one bundle given several others. This approach induces a graphical 
relation among the different bundles.
We will equip these relations with a causal interpretation by additionally
specifying the distribution of $\B{Z}$ under certain interventions on 
the data generating process.

\begin{defi}[Causal graphical models for spatio-temporal processes] \label{defi:CGM}
A causal graphical model for a $p$-dimensional spatio-temporal process $\B{Z}$ 
is a triplet $(\mathcal{S}, \mathcal{G}, \mathcal{P})$ consisting of 
\begin{itemize}
\setlength\itemsep{0em}
\item a family $\mathcal{S} = (S_j)_{j=1}^k$ of non-empty, disjoint sets $S_1, \dots, S_k \subset \{1, \dots, p\}$ 
with $\bigcup_{j=1}^k S_j = \{1, \dots, p\}$,
\item a directed acyclic graph $\mathcal{G}$ with vertices $S_1, \dots, S_k$, and
\item a family $\mathcal{P} = (\mathcal{P}^j)_{j = 1}^k$ of collections $\mathcal{P}^j = \{\P^j_z\}_{z \in \mathcal{Z}_{\card{\emph{PA}_j}}}$
of distributions on $(\mathcal{Z}_{\card{S_j}}, \mathcal{F}_{\card{S_j}})$, where 
for every $j$, 
$\emph{PA}_j := \bigcup_{i: S_i \rightarrow S_j \in \mathcal{G}} S_i$.
Whenever $\emph{PA}_j = \emptyset$, $\mathcal{P}^j$ consists only of a single distribution which we denote by $\P^j$. 
\end{itemize}
Since $\mathcal{G}$ is acyclic, we can without loss of generality assume that $S_1, \dots, S_k$ are indexed such that $S_i \not \to S_j$ 
in $\mathcal{G}$ whenever $i > j$. 
The above components
induce a unique joint distribution $\P$ over $\B{Z}$. For every $F = \bigtimes_{j=1}^k F_j$, it is defined by
\begin{equation} \label{eq:P}
\P(F) = \int_{F_1} \cdots \int_{F_k} \P_{z^{(\emph{PA}_k)}}^k(dz^{(S_k)}) \cdots \P^1(dz^{(S_1)}).
\end{equation}
We call $\P$ the \emph{observational distribution}. 
For each $j \in \{1, \dots, k\}$, the conditional distribution of $\B{Z}^{(S_j)}$ given $\B{Z}^{(\emph{PA}_j)}$ as induced by $\P$
equals 
 $\mathcal{P}^j$. We define an \emph{intervention on} $\B{Z}^{(S_j)}$ as replacing $\mathcal{P}^j$ by another model $\tilde{\mathcal{P}}^j$. This operation results in a new graphical model $(\mathcal{S}, \mathcal{G}, \tilde{\mathcal{P}})$ for $\B{Z}$ which induces, via \eqref{eq:P}, 
a new distribution $\tilde{\P}$, the \emph{interventional distribution}. 
\end{defi}
Assume that we perform an intervention on $\B{Z}^{(S_i)}$. By definition, the resulting interventional distribution 
differs from the observational distribution only in the way in which $\B{Z}^{(S_i)}$ depends on $\B{Z}^{(\text{PA}_i)}$, 
while all other conditional distributions $\B{Z}^{(S_j)} \given \B{Z}^{(\text{PA}_j)}$, $j \not = i$, remain the same. 
This property is analogous to the modularity property of structural causal models 
\citep{haavelmo1944, 
pearl2009causality, peters2017elements}
and justifies a causal interpretation
of the conditionals in $\mathcal{P}$. 
We refer to the graph $\mathcal{G}$ as the \emph{causal structure of} $\B{Z}$, and sometimes write 
$\B{Z} = [\B{Z}^{(S_k)} \given \B{Z}^{(\text{PA}_k)}] \cdots [\B{Z}^{(S_1)}]$ to emphasize this structure.

\subsection{Latent spatial confounder model} \label{sec:LSCM}
Motivated by the example
introduced 
in Section~\ref{sec:conflict}, we are particularly interested in scenarios 
where a target variable $Y$ is causally influenced by a vector of covariates $X$, 
and where $(X,Y)$ are additionally affected by some latent variables $H$. 
In general, 
inferring causal effects under arbitrary influences of latent confounders is impossible, and
we therefore need to impose additional restrictions on the variables in $H$.
We here make the fundamental assumption that they do not vary across time 
(alternatively, one can assume that the hidden variables are invariant over space, 
see Appendix~\ref{sec:exchange_st}).
\begin{defi}[Latent spatial confounder model] \label{defi:LSCM}
Consider a spatio-temporal
process $(\B{X}, \B{Y}, \B{H}) = (X_s^t, Y_s^t, H_s^t)_{(s,t) \in \R^2 \times \N}$ over a real-valued response $Y$, a 
vector of covariates $X \in \R^d$ and a vector of latent variables $H \in \R^\ell$. 
We call a causal graphical model over $(\B{X}, \B{Y}, \B{H})$ with causal structure $[\B{Y} \given \B{X}, \B{H}] [\B{X} \given \B{H}] [\B{H}]$ 
a \emph{latent spatial confounder model (LSCM)} if both of the following conditions hold true for the observational distribution.
\begin{itemize}
\setlength\itemsep{0em}
\item The latent process $\B{H}$ is weakly stationary and time-invariant.
\item There exists a measurable function $f : \R^{d+\ell+1} \to \R$ and
an i.i.d.\ sequence 
$\boldsymbol{\epsilon}^1, \boldsymbol{\epsilon}^2, \dots$ of weakly-stationary spatial error processes, independent of $(\B{X}, \B{H})$, such that
\begin{equation} \label{eq:f}
Y_s^t = f(X_s^t, H_s^t, \epsilon_s^t) \qquad \text{ for all } (s,t) \in \R^2 \times \N.
\end{equation}
We require that for all $x \in \R^d$, $f(x, H_0^1, \epsilon_0^1)$ has finite expectation.
\end{itemize}
\end{defi}
Throughout this section, we assume that $(\B{X}, \B{Y}, \B{H})$ come from an LSCM.
The above definition says that for every $s,t$, $Y_s^t$ depends 
on $(\B{X}, \B{H})$ only via $(X_s^t, H_s^t)$, and that this dependence remains the same for all points in space-time.
Together with the weak stationarity of $\B{H}$ and $\boldsymbol{\epsilon}$, 
this assumption ensures that the average causal effect of $X_s^t$ on $Y_s^t$ (which we introduce below) remains the same for all $s,t$. 
Our model class imposes no restrictions on the marginal distribution of $\B{X}$. 
The spatial dependence structure of the error process $\boldsymbol{\epsilon}$ must have the
same marginal distributions everywhere, but is otherwise unspecified (in particular, $\boldsymbol{\epsilon}$ is not required to be stationary). 
The temporal independence assumption on $\boldsymbol{\epsilon}$ is necessary for our construction 
of resampling tests, see Section~\ref{sec:testing}. 
We now formally define our inferential target. 
\begin{defi}[Average causal effect] \label{defi:fave}
The \emph{average causal effect} of $\B{X}$ on $\B{Y}$ is 
defined as the function $f_{\emph{AVE}(X \to Y)} : \R^d \to \R$, for every $x \in \R^d$ given by%
\begin{equation} \label{eq:fAVE0}
f_{\emph{AVE}(X \to Y)}(x) := \E[f(x, H^1_0, \epsilon_0^1)].
\end{equation}
\end{defi}
Here, 
the causal effect is an average effect in that 
it takes the expectation over both the noise 
variable (as opposed to making counterfactual 
statements \citep{rubin1974estimating}) 
and the hidden variables. 
Alternatively, one may define the inferential 
target in terms of the single realization of $\B{H}$ which manifested itself
in the data, see Appendix~\ref{app:steffen}.%
\footnote{We are grateful to Steffen
Lauritzen for emphasizing this viewpoint.}
The following proposition justifies $f_{\text{AVE}(X \to Y)}$ as a quantification of the causal influence of $\B{X}$ on $\B{Y}$.
\begin{prop}[Causal interpretation] \label{prop:causal_int}
Let $(s,t) \in \R^2 \times \N$ and $x\in \R^d$ be fixed, and consider any intervention on $\B{X}$ such that $X_s^t = x$ holds almost surely in the induced interventional distribution $\P_x$. We then have that 
$
\E_{\P_x} [Y_s^t] 	= f_{\emph{AVE}(X \to Y)}(x).
$
That is, $f_{\emph{AVE}(X \to Y)}(x)$ is the expected value of $Y_s^t$ under any intervention that enforces $X_s^t = x$. 
\end{prop}
In many applications, we do not have explicit knowledge of, or data from, the interventional distributions 
$\P_x$. If we have access to the causal graph, however, we can sometimes compute intervention
effects from the observational distribution. 
In the i.i.d.\ setting,
depending on which variables 
are observed, this can be done by covariate adjustment
or G-computation \citep{pearl2009causality, rubin1974estimating, shpitser2012validity}, for example. 
The following proposition shows a similar result in the case of an LSCM.
It follows directly 
from 
Fubini's theorem.
\begin{figure}[t]
\begin{minipage}{0.24\textwidth}
\begin{center}
\begin{tikzpicture}[xscale=1.2, yscale=1.2, shorten >=1pt, shorten <=1pt]
  \draw (0,1.8) node(title) {\textbf{confounded effect}};
  \draw (0,0.5) node(pconf) {\includegraphics[width=.8\linewidth]{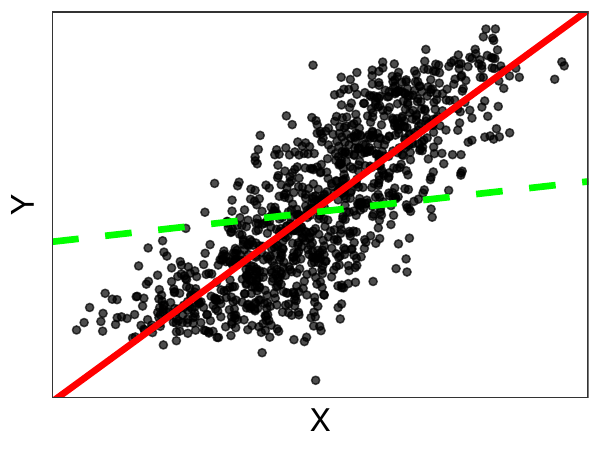}};
  \draw (0,-1.1) node(disag) {spatially disaggr.};
  \draw (1.5,-1) node(tmp) {$ $};
  \draw[-arcsq] (pconf) to [bend right = 30] (tmp);
\end{tikzpicture}
\end{center}
\end{minipage}
\begin{minipage}{0.5\textwidth}
\begin{mdframed}[roundcorner=10pt, innermargin=5pt]
\begin{center}
\begin{tikzpicture}[xscale=1.2, yscale=1.2, shorten >=1pt, shorten <=1pt]
  \draw (0,0) node(pH) {\includegraphics[width=\linewidth]{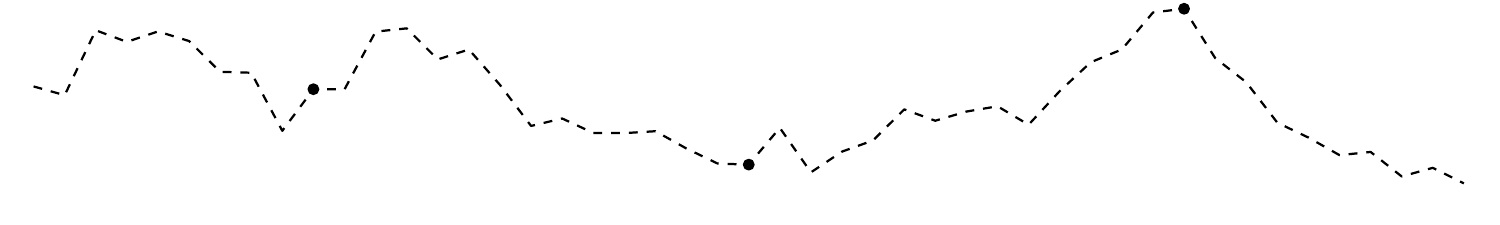}};
  \draw (-2.2,1.5) node(pXY1) {\includegraphics[width=.22\linewidth]{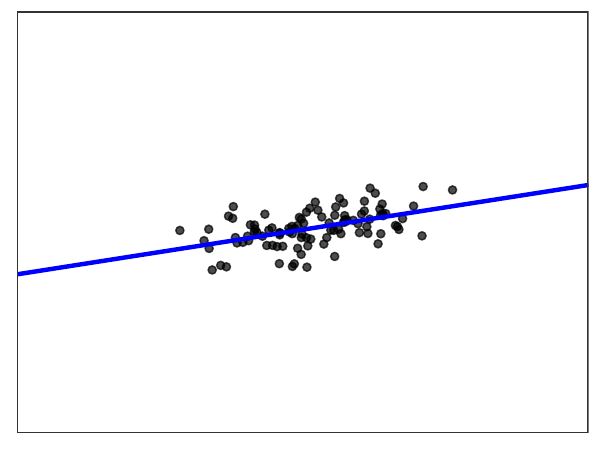}};
  \draw (0,1.5) node(pXY2) {\includegraphics[width=.22\linewidth]{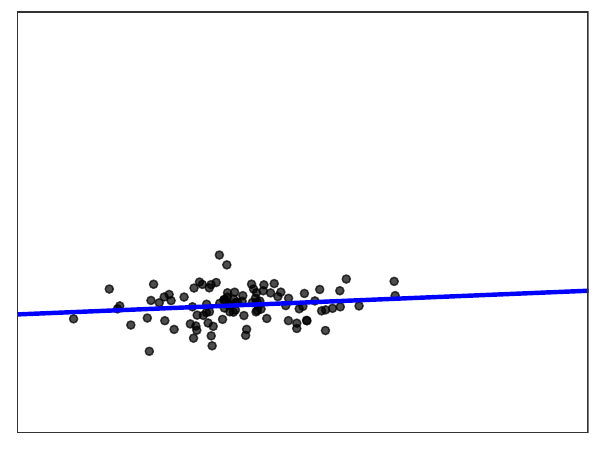}};
  \draw (2.2,1.5) node(pXY3) {\includegraphics[width=.22\linewidth]{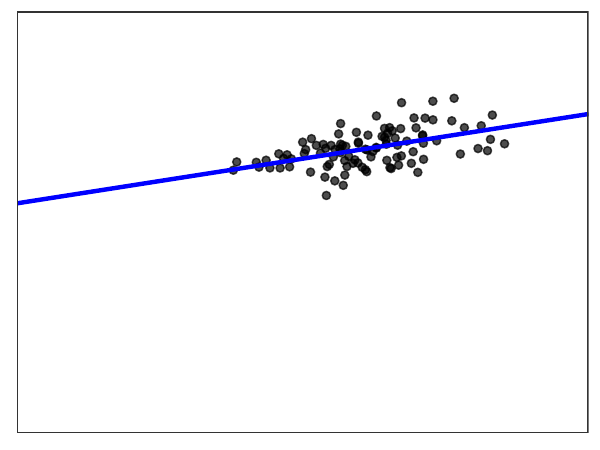}};
  \draw (-1.1,1.5) node {{\footnotesize $\cdots$}};
  \draw (1.1,1.5) node {{\footnotesize $\cdots$}};
  
  \draw (-1.67, 0) node(H1) {};
  \draw (-.02,-.25) node(H2) {};
  \draw (1.67, .36) node(H3) {};
  \draw (-1.55, -.1) node {{\footnotesize $H_{s_1}^1$}};
  \draw (-0.22, 0.05) node {{\footnotesize $H_{s_2}^1$}};
  \draw (1.5, .6) node {{\footnotesize $H_{s_3}^1$}};
  \draw (H1) to [bend left = 20] (pXY1);
  \draw (H2) to [bend right = 10] (pXY2);
  \draw (H3) to [bend right = 30] (pXY3);

\end{tikzpicture}
\end{center}
\end{mdframed}
\end{minipage}
\begin{minipage}{0.24\textwidth}
\begin{center}
\begin{tikzpicture}[xscale=1.2, yscale=1.2, shorten >=1pt, shorten <=1pt]
  \draw (0,1.8) node(title) {\textbf{avg causal effect}};
  \draw (0,0.5) node(pavg) {\includegraphics[width=.8\linewidth]{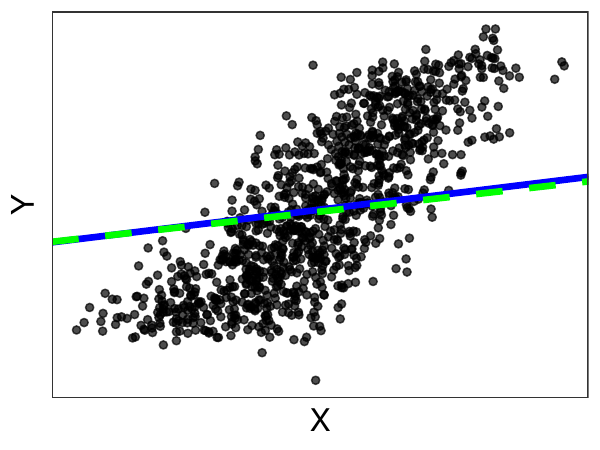}};
  \draw (-1.5,-1) node(tmp) {$ $};
  \draw[-arcsq] (tmp) to [bend right = 30] (pavg);
  \draw (-0.1,-1.1) node(disag) {average};
\end{tikzpicture}
\end{center}
\end{minipage}

\begin{center}
\vspace{-3mm}
for all $s$, 
produce estimate 
$\hat f^m_{Y \vert X}(\B{X}^m_s, \B{Y}^m_s)(\cdot)$
\vspace{-3mm}
\end{center}
\caption{
Conceptual idea for estimating the average causal effect (green line) defined in~\eqref{eq:fAVE0}. 
In both the left and right panel, we do not display 
time and space. 
The middle figure 
shows the data at different locations,
i.e., every small 
plot corresponds to a single time series. 
The dashed curve illustrates 
the latent spatial confounder $H_s^t$ (for visual purposes, we here consider one-dim.\ space).
Due to $H_s^t$, regressing $Y_s^t$ on $X_s^t$ (red line in left plot) leads to a 
biased estimator. %
Our estimator (blue line in right plot) removes this bias.
The procedure is shown in the middle figure:
for all $s$, we observe several instances $(X_s^t, Y_s^t)$, $t = 1, \dots, m$,
with the same conditionals $Y_s^t \given (X_s^t, H^t_s)$ and the same (unobserved) value of $H_s^t$. 
For each realization $h_s$ of $H^1_s$, we can thus estimate the regression $f_{Y \vert (X,H)}(\cdot, h_s)$ 
only using the data $(\B{X}_s^m, \B{Y}_s^m)$ (blue lines in middle figure). 
A final estimate of the average causal effect (blue line in right plot)
is obtained by approximating the expectation in \eqref{eq:fAVEhat} by a sample average over all spatial locations, 
see Section~\ref{sec:estimator}.
}
\label{fig:estimator}
\end{figure}
\begin{prop}[Covariate adjustment] \label{prop:cov_adj}
Let $f_{Y \vert (X,H)}$ denote the regression function $(x,h) \mapsto \E[Y_s^t \given X_s^t = x, H_s^t = h]$ (by definition of an LSCM, this function is the same for all $s,t$). For all $x \in \R^d$, it holds that 
\begin{equation} \label{eq:cov_adj}
f_{\emph{\text{AVE}}(X \to Y)}(x) = \E [f_{Y \vert (X,H)} (x,H_0^1)].
\end{equation}
\end{prop}
Proposition~\ref{prop:cov_adj} shows that $f_{\text{AVE}(X \to Y)}$ is identified from the full observational 
distribution over $(\B{X}, \B{Y}, \B{H})$ (given that the LSCM structure is known). 
Since $\B{H}$ is unobserved, 
the main challenge is to estimate \eqref{eq:cov_adj} merely based on data from $(\B{X},\B{Y})$, see Section~\ref{sec:estimator}.
(We discuss in Appendix~\ref{sec:timevar_conf} 
how to furthermore include observable time-varying covariates.)

\subsection{Estimation of the average causal effect} \label{sec:estimator}
\subsubsection{Definition and consistency}
In practice, we only observe the process $(\B{X},\B{Y})$ at a finite number of points in space and time. 
We assume that at every temporal instance, we 
observe the process at the same spatial locations $s_1, \dots, s_n \in \R^2$ (these locations need not lie on a regular grid).
To simplify notation, 
we furthermore assume that the observed time points 
are $t=1, 2, \ldots, m$,
i.e., we have access to a dataset 
$(\B{X}_n^m, \B{Y}_n^m) = (X_s^t, Y_s^t)_{(s,t) \in \{s_1, \dots, s_n\} \times \{1, \dots, m\}}$. 
The proposed method is based on the following key idea: for every $s \in \{s_1, \dots, s_n\}$, 
we observe several time instances $(X_s^t, Y_s^t)$, $t \in \{1, \dots, m\}$, all with the same 
conditionals $Y_s^t \given (X_s^t, H^t_s)$. Since $\B{H}$ is time-invariant, we can, for every $s$, 
estimate $f_{Y \vert (X,H)}(\cdot, h_s)$ for the (unobserved) realization $h_s$ of $H^1_s$ 
using the data $(X_s^t, Y_s^t)$, $t \in \{1, \dots, m\}$. The expectation in \eqref{eq:cov_adj} 
is then approximated by averaging estimates obtained from different spatial locations. 
This idea is visualized in Figure~\ref{fig:estimator}. 
More formally, our method requires as input a model class for the regressions $f_{Y \vert (X,H)}( \cdot , h)$, 
$h \in \R^\ell$, alongside with a suitable estimator $\hat f_{Y \vert X} = (\hat f^m_{Y \vert X})_{m \in \N}$, 
and returns
\begin{equation} \label{eq:fAVEhat}
\hat f^{nm}_{\text{AVE}(X \rightarrow Y)}(\B{X}_n^m, \B{Y}_n^m)(x) := \frac{1}{n} \sum_{i = 1}^n \hat f^m_{Y \vert X}(\B{X}^m_{s_i}, \B{Y}^m_{s_i})(x),
\end{equation}
an estimator
of the average causal effect \eqref{eq:fAVE0} within the given model class.
In Section~\ref{sec:testing}, we furthermore provide a statistical test for the overall existence of a causal effect.
Our approach may be seen as summarizing the output of a spatially varying regression model \citep[e.g.,][]{gelfand2003spatial}
that is allowed to change arbitrarily from one location to the other (within the model class dictated by $\hat{f}_{Y \vert X}$). 
By permitting such flexibility, our method does not rely on observing data from a continuous or spatially connected domain, 
and accommodates complex influences of the latent variables. 
An implementation can be found in our code package, see Section~\ref{sec:contri}.

To prove consistency of our estimator,
we let the number of observable 
points in space-time increase.
Let therefore
$(s_n)_{n \in \N} \subset \R^2$ be a sequence of spatial coordinates, and consider the array of data 
$(\B{X}_n^m, \B{Y}_n^m)_{n,m\geq 1}$, 
where for every $n,m \in \N$, $(\B{X}_n^m, \B{Y}_n^m) = (X_{s}^t, Y_{s}^t)_{(s,t) \in \{s_1, \dots, s_n\} \times \{1, \dots, m\}}$. 
We want to prove that the corresponding sequence of estimators \eqref{eq:fAVEhat} consistently estimates \eqref{eq:fAVE0}. 
To obtain such a result, we need two central assumptions.

\begin{ass}[LLN for the latent process] \label{ass:LLN}
For all measurable functions $\varphi : \R^\ell \to \R$ with $\E[\card{\varphi(H_0^1)}] < \infty$ it holds 
that $\tfrac{1}{n} \sum_{i=1}^n \varphi(H^1_{s_i}) \to \E[\varphi(H_0^1)]$ in probability as $n \to \infty$.
\end{ass}
Assumption~\ref{ass:LLN} ensures that, for an increasing number of spatial locations at which data are observed, 
the spatial average in~\eqref{eq:fAVEhat} approximates the expectation in~\eqref{eq:fAVE0}. 
The assumption is satisfied, for example, for a stationary Gaussian process $\B{H}^1$ that is sampled regularly
across an increasing spatial domain, see Proposition~\ref{prop:LLN} in Appendix~\ref{sec:suff_cond}.
As the number of observed time points tends to infinity, we furthermore 
require the estimators $\hat{f}^m_{Y \vert X}$
to converge to the integrand in \eqref{eq:fAVE0},
at least in some area $\mathcal{X} \subseteq \R^d$.
\begin{ass}[Consistent estimators of the conditional expectations] \label{ass:fhat}
There exists $\mathcal{X} \subset \R^d$ such that for all $x \in \mathcal{X}$ and $s \in \R^2$, it holds that 
$\hat{f}^m_{Y \vert X}(\B{X}_s^m, \B{Y}_s^m)(x) - f_{Y \vert (X,H)}(x, H_s^1) \to 0$, in probability as  $m \to \infty$.
\end{ass}
A slightly stronger, but maybe more intuitive formulation is to require the above consistency to hold conditionally on $\B{H}$, i.e., 
assuming that for all $x \in \mathcal{X}$, $s \in \R^2$ and almost all $\B{h}$, 
$\hat{f}^m_{Y \vert X}(\B{X}_s^m, \B{Y}_s^m)(x) \to f_{Y \vert (X,H)}(x, h_s^1)$ as $m \to \infty$, in probability under 
$\P(\cdot \given \B{H} = \B{h})$. It follows from the dominated convergence theorem that this assumption implies Assumption~\ref{ass:fhat}. 
Under Assumptions~\ref{ass:LLN}~and~\ref{ass:fhat}, 
we obtain the following consistency result. %
\begin{thm}[Consistent estimator of the average causal effect] \label{thm:consistency}
Let $(\B{X}, \B{Y}, \B{H})$ come from an LSCM as defined in Definition~\ref{defi:LSCM}.
Let $(s_n)_{n \in \N}$ be a sequence of spatial coordinates such that the 
marginalized process $(H^1_{s_n})_{n \in \N}$ satisfies
Assumption~\ref{ass:LLN}, 
and assume that for all $x \in \mathcal{X}$,
$\E[\card{f_{Y \vert (X,H)}(x, H_0^1)}] < \infty$. 
Let furthermore $\hat f_{Y \vert X} = (\hat f^m_{Y \vert X})_{m \in \N}$ be an estimator 
satisfying Assumption~\ref{ass:fhat}. We then have the following consistency result. 
For all $x \in \mathcal{X}$, $\delta > 0$ and $\alpha > 0$, there exists $N \in \N$ 
such that for all $n \geq N$ we can find $M_n \in \N$ such that for all 
$m \geq M_n$ 
we have that 
\begin{equation} \label{eq:Plim}
\P \left( \left \vert \hat f^{n m}_{\emph{AVE}(X \rightarrow Y)}(\B{X}_{n}^{m}, \B{Y}_{n}^{m})(x) -  f_{\emph{AVE}(X \rightarrow Y)}(x) \right \vert > \delta \right) \leq \alpha.
\end{equation}
\end{thm}

Apart from the LSCM structure, the above result does not rely on any particular distributional 
properties of the data. 
Assumptions~\ref{ass:LLN}~and~\ref{ass:fhat}
do not impose strong restrictions on the 
data generating process and hold true for several model classes, 
including linear and nonlinear models. 
In Appendix~\ref{sec:suff_cond}, we 
provide sufficient conditions under which these assumptions are true.

\subsubsection{An example LSCM}
\begin{figure}[t]
	\centering
	\begin{minipage}{0.28\linewidth}
		\begin{center}
		{\bf $\bar{\B{H}}$ \hspace{3.2mm}   $\tilde{\B{H}}$ \hspace{3.2mm}   $\B{X}$ \hspace{3.2mm}   $\B{Y}$} 
		
		\vspace{1mm}
		
		\includegraphics[width=.9\textwidth]{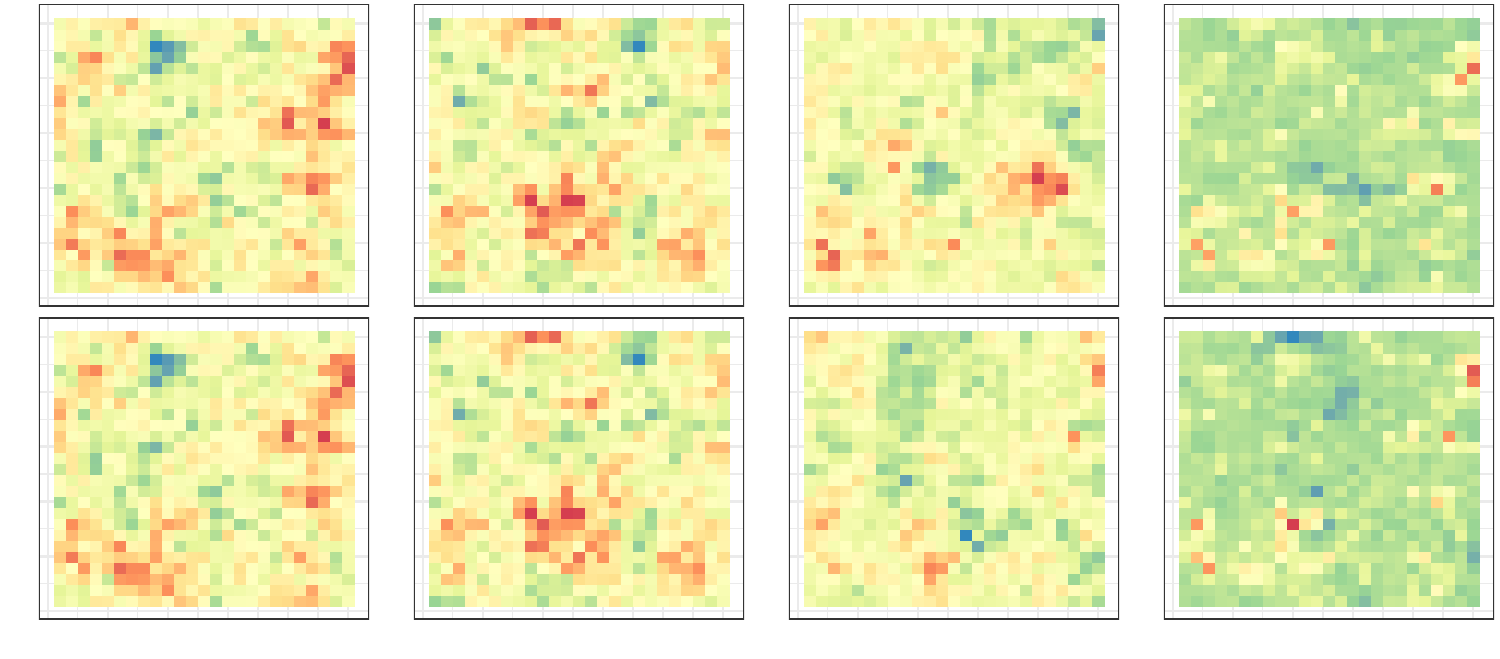}

		\vspace{-4mm}

		{\bf \, \, $\vdots$ \hspace{3.2mm}  \, $\vdots$ \hspace{3.2mm}  \, $\vdots$ \hspace{3.2mm}  \, $\vdots$}

		\vspace{4mm}

		\includegraphics[width=.9\textwidth]{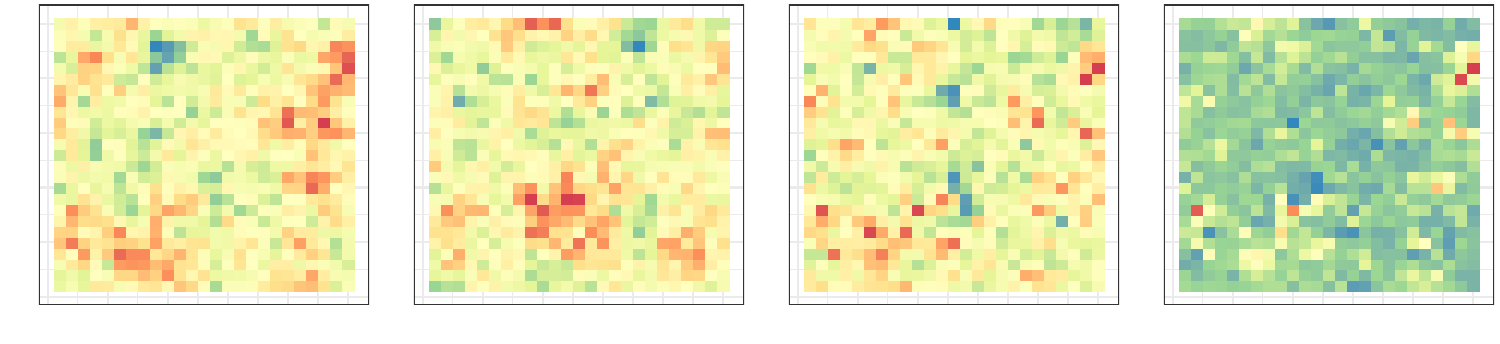}
		\end{center}
	\end{minipage}
	\begin{minipage}{0.3\linewidth}
		\begin{center}
		$ $
		
		\vspace{7mm}
		
		\includegraphics[width=.92\textwidth]{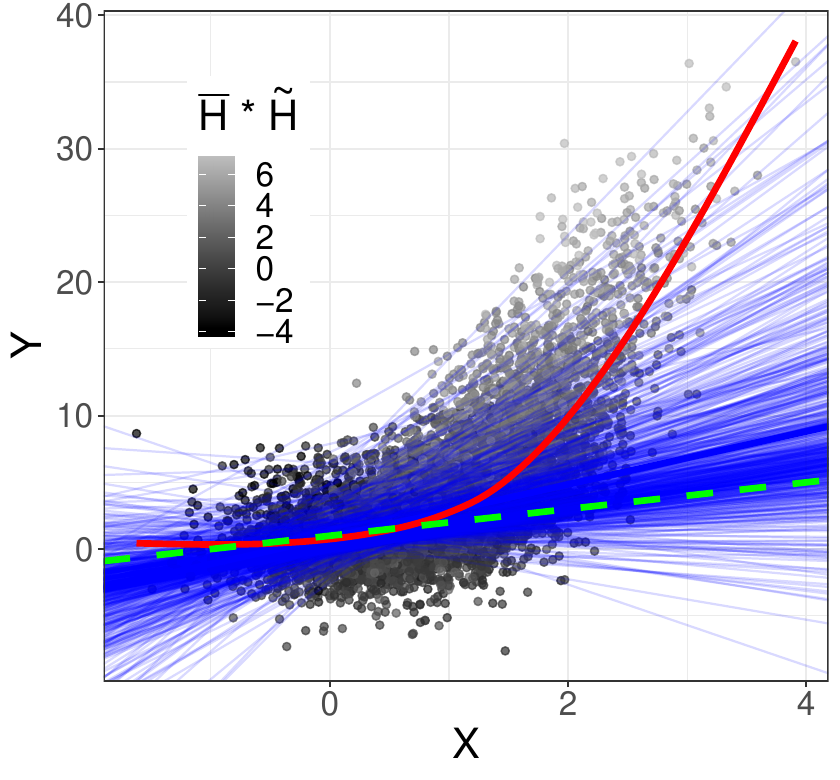}
		\end{center}
	\end{minipage}
	\begin{minipage}{0.37\linewidth}
		\begin{center}
		$ $
		
		\vspace{7mm}
		
		\includegraphics[width=.9\textwidth]{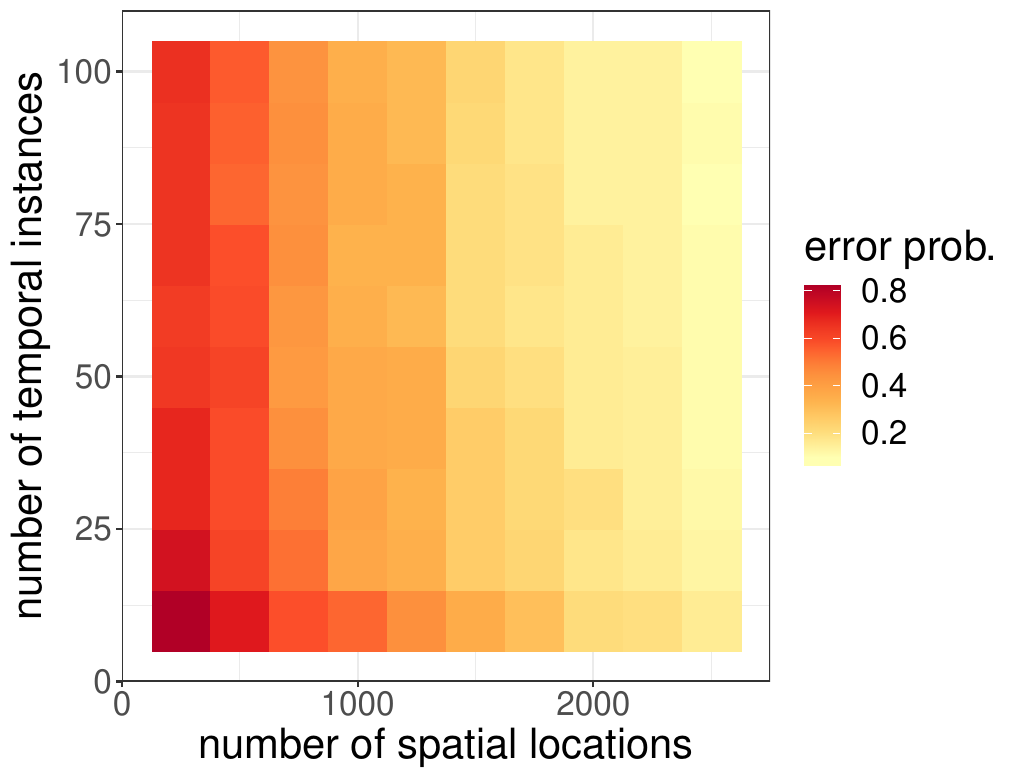}
		\end{center}
	\end{minipage}
	\caption{Results for applying our methodology to 
	Example~\ref{ex:sim}. 
	The left panel shows a sample dataset of $(\B{X}, \B{Y}, \B{H})$ observed 
	at the spatial grid $\{1, \dots, 25\}^2$ and at several 
	time instances. 
	The middle panel illustrates 
	our method applied to the same dataset. The average causal effect, our inferential
	target, is indicated by a 
	green line. Due to confounding by $H_s^t$, a standard 
	nonlinear regression (red curve) severely 
	overestimates the causal influence of $X_s^t$ on $Y_s^t$.
	By regressing $Y_s^t$ on $X_s^t$ in each 
	location separately
	(thin blue lines), and aggregating the results into a final estimate (thick blue line), 
	all spatial
	confounding is removed. In the right panel, we investigate the consistency result from 
	Theorem~\ref{thm:consistency} empirically. 
	For increasing $n$ and $m$, 
	we generate 
	several datasets $(\B{X}_{n,i}^m, \B{Y}_{n,i}^m)$, $i = 1, \dots, 100$, compute estimates $\hat{\beta}^{nm}_i$
	of the causal coefficients $\beta$, and use these to compute empirical 
	error probabilities $\hat{\P} (\norm{\hat{\beta}^{nm} - \beta}_2 > \delta)$. 
	In the above plot, we have chosen $\delta = 0.2$. As $n$ and $m$ increase, the error probability
	tends towards zero. 
	}
	\label{fig:example}
\end{figure}

To illustrate the consistency result in Theorem~\ref{thm:consistency}, we now consider a simple example with one covariate ($d = 1$) and
two hidden variables ($\ell = 2$).
\begin{ex}[Latent Gaussian process and a linear average causal effect] \label{ex:sim}
Let $\boldsymbol{\zeta}, \boldsymbol{\psi}, \boldsymbol{\xi}^{t}, \boldsymbol{\epsilon}^{t}$, 
$t \in \N$, be independent versions
of a univariate stationary spatial
Gaussian process with mean $0$ and covariance function $u \mapsto  e^{-\tfrac{1}{2} \norm{u}_2}$. 
For notational simplicity, let $\bar{\B{H}}$ and $\tilde{\B{H}}$ denote the respective first and second coordinate process 
of $\B{H}$. We define a marginal distribution over $\B{H}$ and conditional distributions $\B{X} \given \B{H}$ and 
$\B{Y} \given (\B{X}, \B{H})$ by specifying that for all $(s,t) \in \R^2 \times \N$,
\begin{align*}
H_s^t &= (\bar{H}_s^t, \tilde{H}_s^t) = (\zeta_s, 1 + \tfrac{1}{2} \zeta_s + \tfrac{\sqrt{3}}{2} \psi_s), \\
X_s^t &= \exp(-\norm{s}_2^2/1000) + (0.2 + 0.1 \cdot \sin( 2 \pi t / 100)) \cdot \bar{H}_s^t \cdot \tilde{H}_s^t + 0.5 \cdot \xi_s^t, \\
Y_s^t &= (1.5 + \bar{H}_s^t \cdot \tilde{H}_s^t) \cdot X_s^t+ (\bar{H}_s^t)^2 + \card{\tilde{H}_s^t} \cdot \epsilon_s^t.
\end{align*}
Interventions on $\B{X}$, $\B{Y}$ or $\B{H}$ are defined as in Definition~\ref{defi:CGM}.
In this LSCM, the average causal effect $f_{\emph{AVE}(X \to Y)}$ is the linear function $x \mapsto \beta_0 + \beta_1 x$, 
with $\beta_0 := \E[(\bar{H}_0^1)^2] = 1$ and $\beta_1 := 1.5 + \E[\bar{H}_0^1 \cdot \tilde{H}_0^1] = 2$. 
We define a spatial sampling scheme $(s_i)_{i \in \N}$ 
for every $j \in \N$ and $k \in \{1, \dots, 25\}$ by 
$s_{25 \cdot (j-1) + k} = (j,k)$. 
Given a sample $(\B{X}_n^m, \B{Y}_n^m) = (X_s^t, Y_s^t)_{(s,t) \in \{s_1, \dots, s_n\} \times \{1, \dots, m\}}$ from $(\B{X}, \B{Y})$, 
we construct an estimator of $f_{\emph{AVE}(X \to Y)}$ by
\begin{equation} \label{eq:beta0hat}
\hat f_{\emph{AVE}(X \to Y)}^{nm}(\B{X}_n^m, \B{Y}_n^m)(x) = \frac{1}{n} \sum_{i = 1}^n (1 \ x) \, \hat \beta_{\emph{OLS}}^m(\B{X}_{s_i}^m, \B{Y}_{s_i}^m),
\end{equation}
where $\hat \beta_{\emph{OLS}}^m(\B{X}_{s_i}^m, \B{Y}_{s_i}^m) \in \R^{2}$ is the OLS estimator for the linear regression 
at spatial location $s_i$, that is 
of 
$\B{Y}_{s_i}^m = (Y_{s_i}^1, \ldots, Y_{s_i}^m)$
on 
$\B{X}_{s_i}^m = (X_{s_i}^1, \ldots, X_{s_i}^m)$
(we assume that the regression includes an intercept term). 
It follows by Propositions~\ref{prop:LLN}~and~\ref{prop:fhat} (see in particular Example~\ref{ex:mixing} and Remark~\ref{rem:mixing2} in Appendix~\ref{app:examples}) that 
Assumptions~\ref{ass:LLN}~and~\ref{ass:fhat} are satisfied.\footnote{Strictly speaking, Example~\ref{ex:mixing} and Remark~\ref{rem:mixing2} show that (L1)--(L3) are satisfied for bounded basis functions. We are confident that 
the same holds true in the current example.}
Hence, \eqref{eq:beta0hat} is a consistent estimator of $f_{\emph{AVE}(X \to Y)}$. 
\end{ex}
Figure~\ref{fig:example} illustrates our procedure using a numerical experiment based on Example~\ref{ex:sim}.%
\footnote{In this example, a standard regression approach overestimates the causal effect. Similarly,
one can construct examples where the causal effect would be underestimated.}
The example shows that we can estimate causal effects 
even under complex influences of the latent process $\B{H}$. 
To construct the estimator $\hat{f}_{\text{AVE}(X \to Y)}^{nm}$, we have used 
that the 
influence of $(\B{X}, \B{H})$ on $\B{Y}$ is linear in $\B{X}$. 
However, we do not assume knowledge 
of the particular functional dependence of $\B{Y}$ on $\B{H}$; we obtain consistency under any 
influence of the form $Y_s^t = f_1(H_s^t) \cdot X_s^t + f_2(H_s^t, \epsilon_s^t)$, see Proposition~\ref{prop:fhat}.

\subsubsection{Extensions}
Our methodology can be extended in several directions. 
In Appendix~\ref{sec:extensions}, 
we provide details on how to model the influence of observed (time- and space-varying) 
confounders and temporally lagged causal effects of $\B{X}$ on $\B{Y}$. We furthermore argue that our method also 
applies in situations where the roles of time and space are interchanged, i.e., when the 
hidden confounders are time-varying, but 
remain static across space. 
The extension which includes observed confounders is furthermore explored empirically in Appendix~\ref{sec:timevar_conf_sim}.

\section{Testing for the existence of causal effects} \label{sec:testing}

The previous section has been concerned with the quantification and estimation of the causal effect of $X$ on $Y$. 
In this section, we introduce hypothesis tests 
for this effect being zero.
We consider the null hypothesis
\begin{equation*}
H_0: (\B{X}, \B{Y}) \text{ come from an LSCM with a function } f \text{ that is constant w.r.t.\ } X_s^t,
\end{equation*}
which formalizes the assumption of ``no causal effect of $X$ on $Y$'' within the LSCM framework.
We construct a 
hypothesis test for $H_0$ using data resampling. 
Our approach acknowledges the existence of spatial dependence in the data without modeling it explicitly. 
It thus does not rely on distributional assumptions apart from the LSCM structure.

For the construction of a resampling test, we closely follow the setup presented in \cite{pfister2018kernel}. 
We require a data permutation scheme which, 
under the null hypothesis, leaves the distribution of the data unaffected. In particular, 
it must preserve the dependence between $\B{X}$ and $\B{Y}$ that is induced by $\B{H}$.
The idea is to permute observations of $\B{Y}$ corresponding to the same (unobserved) 
values of $\B{H}$. Since $\B{H}$ is assumed to be constant within every spatial location, 
this is achieved by permuting $\B{Y}$ along the time axis. 
Let $(\B{X}_n^m, \B{Y}_n^m)$ be the observed data. 
For every $(\B{x},\B{y}) \in \R^{(d+1)\times n \times m}$ and every permutation $\sigma$ of the elements in $\{1, \dots, m\}$, 
let $\sigma(\B{x},\B{y}) \in \R^{(d+1)\times n \times m}$ denote the permuted array with entries $(\sigma(x,y))^t_{s} = (x_s^t, y_s^{\sigma(t)})$. We then have the following exchangeability property. 
\begin{prop}[Exchangeability] \label{prop:exchange}
For any permutation $\sigma$ of the elements in $\{1, \dots, m\}$, we have that, under $H_0$,
$\sigma(\B{X}_n^m, \B{Y}_n^m)$ is equal in distribution to $(\B{X}_n^m, \B{Y}_n^m)$.
\end{prop}
Proposition~\ref{prop:exchange} is the cornerstone for the construction of a valid resampling test. 
Under the null hypothesis, we can compute pseudo-replications of the observed sample $(\B{X}_n^m, \B{Y}_n^m)$
using the permutation scheme described above. 
Given any 
test statistic $\hat T : \R^{(d+1) \times n \times m} \to \R$, 
we obtain a $p$-value for $H_0$ by comparing the value of $\hat T$ calculated on the original dataset with the 
empirical null distribution of $\hat T$ obtained from the resampled datasets. 
The choice of $\hat T$ determines the power of the test.
More formally, 
let $M := m!$ and let $\sigma_1, \dots, \sigma_M$ be all permutations of the elements in $\{1, \dots, m \}$. By Proposition~\ref{prop:exchange}, each $\sigma_i$ 
yields 
a new dataset with the same distribution as $(\B{X}_n^m, \B{Y}_n^m)$.
Let $B \in \N$ and let $k_1, \dots, k_B$ be independent, uniform draws from $\{1, \dots, M\}$. 
For every $(\B{x}, \B{y})$, we define
\begin{equation*} 
p_{ \hat T}(\B{x}, \B{y}) := \frac{1 + \card{\{b \in \{1, \dots, B \} : \hat T(\sigma_{k_b}(\B{x}, \B{y})) \geq \hat T(\B{x}, \B{y}) \}}}{1+B},
\end{equation*}
and construct for every $\alpha \in (0,1)$ a test $\varphi_{ \hat T}^\alpha : \R^{(d+1) \times n \times m} \to \{0,1\}$ of $H_0$ defined by $\varphi^\alpha_{\hat T} = 1 :\Leftrightarrow p_{ \hat T} \leq \alpha$.\footnote{Two-sided tests can be obtained using $p_{\hat T, 2 \text{-sided}} := \min (1, 2 \cdot \min (p_{ \hat T}, p_{- \hat T}))$, for example.}
The following level guarantee for $\varphi_{ \hat T}^\alpha$ follows directly from \cite[Proposition~B.4]{pfister2018kernel}. 
\begin{cor}[Level guarantee of resampling test] \label{cor:level}
Assume that for all $k, \ell \in \{1, \dots, B\}$, $k \not = \ell$, it holds that, under $H_0$,
$
\P(\hat T(\sigma_{k}(\B{X}_n^m, \B{Y}_n^m)) = \hat T(\sigma_{\ell}(\B{X}_n^m, \B{Y}_n^m))) = 0
$.
Then, for every $\alpha \in (0, 1)$, the test $\varphi_{ \hat T}^\alpha$ has correct level $\alpha$.
\end{cor}
Corollary~\ref{cor:level} ensures valid test level for a large class of test statistics. 
The particular choice of test statistic should depend on the alternative hypothesis
that we seek to have power against. Within the LSCM model class, it makes sense to 
quantify deviations from the null hypothesis using functionals of the average causal 
effect, i.e., $T = \psi (\fave)$ 
for some suitable function $\psi$. As a test statistic, we then use the plug-in estimator 
$
\hat{T} (\B{X}_n^m, \B{Y}_n^m) = \psi (\favehat  (\B{X}_n^m, \B{Y}_n^m)).
$
An implementation of the above testing procedure is contained in our code package, see Section~\ref{sec:contri}.
A power-analysis can be found in Appendix~\ref{sec:power}.

\section{Conflict and forest loss in Colombia} \label{sec:conflict_cont}
We return to the problem of conflict ($\B{X}$) and forest loss ($\B{Y}$). %
As discussed in Section~\ref{sec:conflict}, there may exist several confounders of the (potential) causal 
effect of $\B{X}$ on $\B{Y}$. We believe that 
many of these vary at a relatively
low temporal frequency (e.g., population density, market infrastructure), justifying the application
of the methods developed above.

\subsection{Data description and preprocessing} \label{sec:data}

Our analysis is based on two main datasets: (1) a remote sensing-based forest loss 
dataset for the period 2000--2018, which identifies annual forest loss at a spatial resolution 
of $30 \si{\m} \times 30 \si{\m}$ using Landsat satellites \citep{hansen2013high}. 
Here, forest loss is defined as complete canopy removal. 
(2) Spatially explicit information on conflict events from 2000 to 2018, based on the
Georeferenced Event Dataset (GED) from the Uppsala Conflict Data Program (UCDP)
\citep{croicu2015ucdp}. In this dataset, a conflict event is defined as ``an incident 
where armed force was used by an organized actor against another organized actor, or 
against civilians, resulting in at least one direct death at a specic location and a specifc 
date". Such events were identified through global newswire 
reporting, global monitoring of local news, and other secondary sources such as reports by 
non-governmental organizations (for information on the data collection as well as control for 
quality and consistency of the data, please refer to \cite{croicu2015ucdp}). We homogenized these datasets through aggregation to a spatial 
resolution of $10 \si{\km} \times 10 \si{\km}$ by averaging the annual forest loss
within each grid, and by counting all conflict events occurring in the same year and within
the same grid. 
As a proxy for accessibility, we additionally calculated, for each spatial grid, the 
average Euclidean distance to the closest road segment,
using spatial data from \url{https://diva-gis.org} containing all primary and secondary 
roads in Colombia. We regard this variable as relatively constant throughout the considered time-span.

\subsection{Quantifying the causal influence of conflict on forest loss} \label{sec:conflict_quant}
We assume that $(\B{X}, \B{Y})$ come from an LSCM as defined in Defintion~\ref{defi:LSCM}.
Since $X_s^t$ is binary, we can characterize the causal 
influence of $\B{X}$ on $\B{Y}$ by $T := f_{\text{AVE}(X \to Y)}(1) - f_{\text{AVE}(X \to Y)}(0)$, i.e.,
the difference in expected forest loss $\E[Y_s^t]$ under the respective interventions enforcing 
conflict ($X_s^t := 1$) and peace ($X_s^t := 0$). 
Positive values of $T$ correspond to an augmenting effect of conflict on forest 
loss and negative values correspond to a reducing effect. 
Our goal is to estimate $T$, and to test the hypothesis $H_0: T = 0$ (no causal effect of $\B{X}$ on $\B{Y}$). 
To construct an estimator of the average causal effect of the form \eqref{eq:fAVEhat}, 
we require estimators of the conditional expectations $x \mapsto f_{Y \vert (X,H)}(x,h)$. Since $X_s^t$
is binary, we use simple sample averages of the response variable. To make the resulting estimator of $f_{\text{AVE}(X \to Y)}$
well-defined, we omit all locations which do not contain at least one observation from each of 
the regimes $X_s^t = 0$ and $X_s^t = 1$. More precisely, let $(\B{X}_n^m, \B{Y}_n^m)$ be the observed dataset. 
We then use the estimator
\begin{equation} \label{eq:fAVEhat01}
\hat{f}^{nm}_{\text{AVE}(X \to Y)}(\B{X}_n^m, \B{Y}_n^m)(x) = \frac{1}{\card{\mathcal{I}_n^m}} \sum_{i \in \mathcal{I}_n^m} \frac{1}{\card{\{t : X_{s_i}^t = x \}}} \sum_{t : X_{s_i}^t = x} Y_{s_i}^t, \qquad x \in \{0,1\},
\end{equation}
where $\mathcal{I}_n^m := \{i \in \{1, \dots, n\} \st \exists \ t_0, t_1 \in \{1, \dots, m\} \text{ such that } X_{s_i}^{t_0} = 0 \text{ and }  X_{s_i}^{t_1} = 1\}$. 
To test $H_0$, we use the resampling test from Section~\ref{sec:testing} with
$\hat{T} = \favehat(1) - \favehat(0)$. In Appendix~\ref{sec:bias}, we argue that, 
under additional assumptions on the underlying LSCM, the above estimator is approximately unbiased.
Note, however, 
that even if the above estimator is biased, this does not affect the level 
guarantee of our testing procedure---we obtain a valid test level for any test statistic $\hat{T}$ satisfying
the assumptions of Corollary~\ref{cor:level}.

\subsection{Alternative assumptions on the causal structure} \label{sec:conflict_comparison}

To emphasize the relevance of the assumed causal structure, we compare our method with two alternative approaches 
based on different assumptions about the 
ground truth: Model~1 assumes no confounders of $(\B{X}, \B{Y})$ and Model~2 assumes that the only confounder is the observed
process $\B{W}$ (mean distance to a road).
In reality, we expect the existence of several confounders in addition to $\B{W}$, e.g., population density or market infrastructure.
Even though none of the models may be a precise description 
of the data generating mechanism, 
we therefore regard both Models~1~and~2 
as less realistic than the LSCM. 
In both models we can, similarly to Definition~\ref{defi:fave}, define the average causal effect of $\B{X}$ on $\B{Y}$. 
Under Model~1, $f_{\text{AVE}(X \to Y)}$ coincides with the conditional expectation of $Y_s^t$ 
given $X_s^t$, which can be estimated simply using sample averages (as is done in Figure~\ref{fig:col_fl} left). Under Model~2, 
$f_{\text{AVE}(X \to Y)}$ can, analogously to Proposition~\ref{prop:cov_adj}, be computed by adjusting for the confounder $\B{W}$.
For each $x \in \{0,1\}$, we obtain an estimate $\hat f^{nm}_{\text{AVE}(X \to Y)}(x)$ by calculating sample averages of $\B{Y}$ 
across different subsets $\{(s,t) \in \R^2 \times \N \st X_s^t = x, \, W_s^t \in \mathcal{W}_j \}, j \in \mathcal{J}$ 
(we here construct these by considering 100 equidistant quantiles of $\B{W}$), and subsequently 
averaging over the resulting values. %
In both models, we furthermore test the hypothesis of no causal effect of $\B{X}$ on $\B{Y}$ using approaches similar to 
the ones presented in
Section~\ref{sec:testing}.
Under the LSCM assumption, we have constructed a permutation scheme that permutes the values of $\B{Y}$ along the time axis, to preserve 
the dependence between $\B{Y}$ and 
$\B{H}$, see Proposition~\ref{prop:exchange}.
Similarly, we construct a permutation scheme for Model~2
by permuting observations of $\B{Y}$ corresponding to similar values of the confounder $\B{W}$ (i.e., values within the same quantile range). 
Under the null hypothesis corresponding 
to Model~1, $\B{X}$ and $\B{Y}$ are (unconditionally) independent, and we therefore permute the values of $\B{Y}$ completely at random.
Strictly speaking, the permutation schemes for Models~1~and~2 require additional exchangeability assumptions on $\B{Y}$ in order to 
yield valid tests. %
In Appendix~\ref{app:permutation}, we repeat the analysis for Model~1 using a spatial block-permutation to account 
for the spatial
dependence in $\B{Y}$, and obtain similar results.

\subsection{Results} \label{sec:conflict_results}
\begin{figure}
\begin{center}
\begin{minipage}{0.05\linewidth}
\centering
$H_0:$
\end{minipage}
\begin{minipage}{0.3\linewidth}
\centering
{\bf Model 1} \vspace{4mm}

\begin{tikzpicture}[xscale=1.2, yscale=1.2, shorten >=1pt, shorten <=1pt]  
  \draw (0,0) node(x) [observedsmall] {$\B{X}$};
  \draw (1.2,0) node(y) [observedsmall] {$\B{Y}$};
  \draw (0.6,0.75) node(z) [observedsmall, draw=white] {\color{white} $\B{W}$};
  \draw[-arcsq, thick, green!70!black] (x) to (y);
\end{tikzpicture}

\vspace{4mm}

\begin{tikzpicture}[xscale=1.2, yscale=1.2, shorten >=1pt, shorten <=1pt]  
  \draw (0,0) node(x) [observedsmall] {$\B{X}$};
  \draw (1.2,0) node(y) [observedsmall] {$\B{Y}$};
  \draw (0.6,0.75) node(z) [observedsmall, draw=white] {\color{white} $\B{W}$};
\end{tikzpicture}

\vspace{4mm}

\includegraphics[width=.9\linewidth]{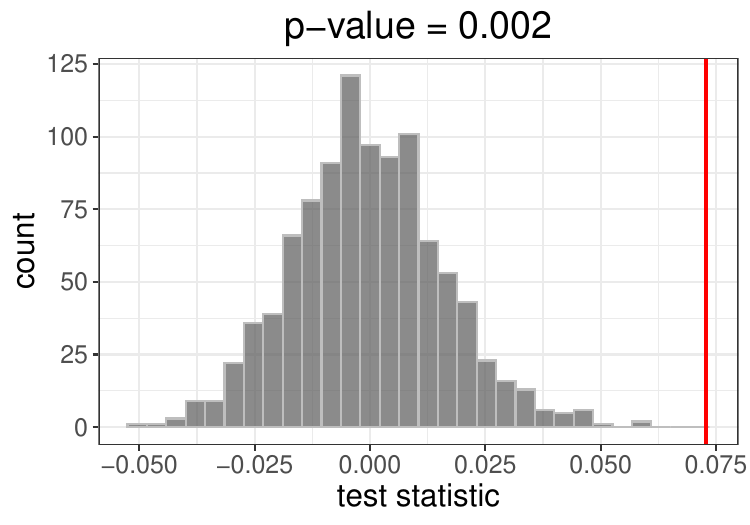}
\end{minipage}
\begin{minipage}{0.3\linewidth}
\centering
{\bf Model 2} \vspace{4mm}

\begin{tikzpicture}[xscale=1.2, yscale=1.2, shorten >=1pt, shorten <=1pt]  
  \draw (0,0) node(x) [observedsmall] {$\B{X}$};
  \draw (1.2,0) node(y) [observedsmall] {$\B{Y}$};
  \draw (0.6,0.75) node(z) [observedsmall] {$\B{W}$};
  \draw[-arcsq, thick, green!70!black] (x) to (y);
  \draw[-arcsq] (z) to (y);
  \draw[-arcsq] (z) to (x);
\end{tikzpicture}

\vspace{4mm}

\begin{tikzpicture}[xscale=1.2, yscale=1.2, shorten >=1pt, shorten <=1pt]  
  \draw (0,0) node(x) [observedsmall] {$\B{X}$};
  \draw (1.2,0) node(y) [observedsmall] {$\B{Y}$};
  \draw (0.6,0.75) node(z) [observedsmall] {$\B{W}$};
  \draw[-arcsq] (z) to (y);
  \draw[-arcsq] (z) to (x);
\end{tikzpicture}

\vspace{4mm}

\includegraphics[width=.9\linewidth]{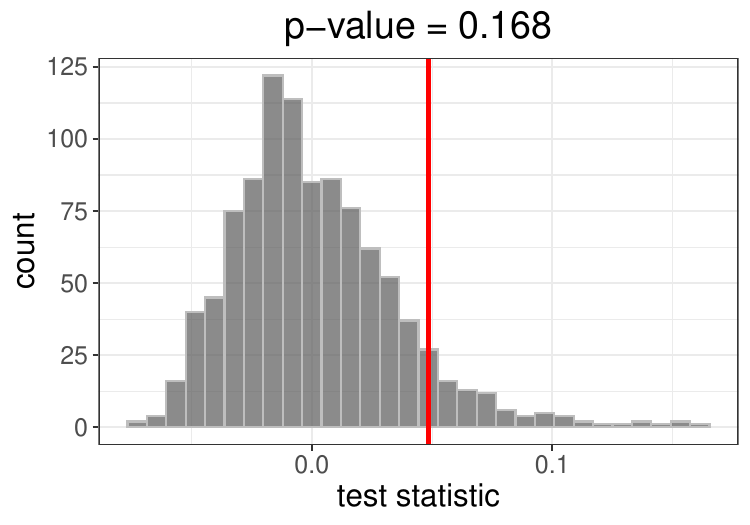}
\end{minipage}
\begin{minipage}{0.3\linewidth}
\centering
{\bf LSCM} \vspace{4mm}

\begin{tikzpicture}[xscale=1.2, yscale=1.2, shorten >=1pt, shorten <=1pt]  
  \draw (0,0) node(x) [observedsmall] {$\B{X}$};
  \draw (1.2,0) node(y) [observedsmall] {$\B{Y}$};
  \draw (0.6,0.75) node(h) [unobservedsmall] {$\B{H}$};
  \draw[-arcsq, thick, green!70!black] (x) to (y);
  \draw[-arcsq, dashed] (h) to (y);
  \draw[-arcsq, dashed] (h) to (x);
\end{tikzpicture}

\vspace{4mm}

\begin{tikzpicture}[xscale=1.2, yscale=1.2, shorten >=1pt, shorten <=1pt]  
  \draw (0,0) node(x) [observedsmall] {$\B{X}$};
  \draw (1.2,0) node(y) [observedsmall] {$\B{Y}$};
  \draw (0.6,0.75) node(h) [unobservedsmall] {$\B{H}$};
  \draw[-arcsq, dashed] (h) to (y);
  \draw[-arcsq, dashed] (h) to (x);
\end{tikzpicture} 

\vspace{4mm}

\includegraphics[width=.9\linewidth]{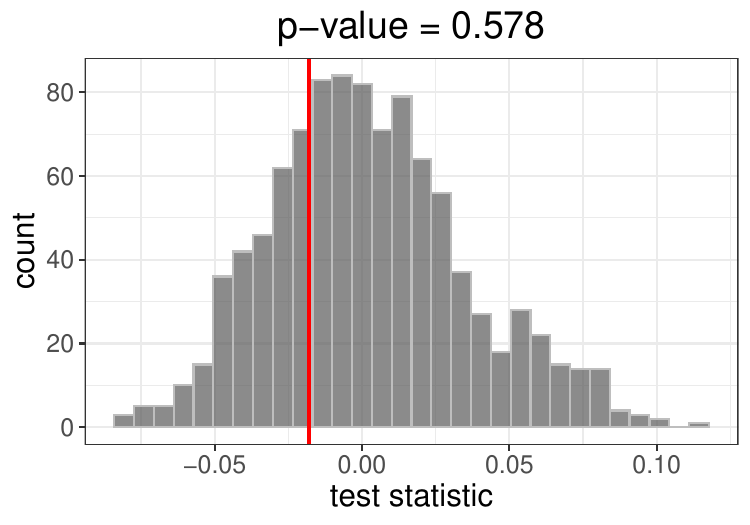}
\end{minipage}
\end{center}
\caption{
Testing for a causal influence of conflict ($\B{X}$) on forest loss ($\B{Y}$) using our method (right)
and two alternative approaches (left and middle) which are based on different 
and arguably less realistic
assumptions on the causal structure.
The process $\B{W}$ corresponds to the mean distance to a road, and $\B{H}$ represents unobserved time-invariant confounders. 
Each of the above models gives rise to a different expression for the test statistic 
$\hat T = \hat{f}^{nm}_{\text{AVE}(X \to Y)}(1) - \hat{f}^{nm}_{\text{AVE}(X \to Y)}(0)$ (indicated by red vertical bars), 
see Sections~\ref{sec:conflict_quant}~and~\ref{sec:conflict_comparison}.
The gray histograms illustrate the empirical null distributions of $\hat T$ under the respective null hypotheses obtained from 999 resampled datasets. 
The results show that our conclusions about the causal influence of conflict on forest loss strongly 
depend on the assumed causal structure: under Model~1, there is a positive, highly significant effect ($\hat{T} = 0.073$). When adjusting for the
confounder $\B{W}$, the effect size decreases and becomes insignificant ($\hat{T} = 0.049$). %
When applying our proposed methodology, the estimated effect is negative ($\hat{T} = - 0.018$). 
}
\label{fig:tests}
\end{figure}

The results of applying our method and the two alternative approaches to the entire study region are depicted in Figure~\ref{fig:tests}. 
Under Model~1, there is an enhancing, highly significant causal effect of conflict on forest loss 
($\hat T = 0.073$, $P = 0.002$). When adjusting for accessibility (quantified by $\B{W}$, Model~2), 
the size of the estimated causal effect shrinks, and becomes insignificant ($\hat T = 0.049$, $P = 0.168$).
(Note that  we have considered other confounders, too, yet obtained similar results.
For example, when adjusting for population density, which we consider as moderately 
temporally varying, we obtain $\hat{T} = 0.038$ and $P = 0.214$.) 
When applying the 
methodology proposed in this paper, that is, adjusting for all time-invariant confounders, 
the estimated effect swaps sign ($\hat T = -0.018$, $P = 0.578$), but is insignificant.
We also tested for spatial spill-over effects by modeling causal effects that are spatially 
lagged by up to one or two pixels, respectively. In both cases, the effect was negative and insignificant (results not shown here).
One reason for the above non-finding could be the time delay between the proposed cause (conflict) and effect (forest loss). 
To account for this potential issue, we also test for a causal effect of $\B{X}$ on $\B{Y}$ that is temporally lagged by one year, i.e.,
we use an estimator similar to \eqref{eq:fAVEhat01}, where we compare the average forest loss 
succeeding conflict events with the average forest loss succeeding non-conflict events. 
Again, the estimated influence of $\B{X}$ on $\B{Y}$ is negative and insignificant ($\hat T = -0.0293$, $P = 0.354$). 
Additionally, we perform alternative versions of the last two tests to account for potential autocorrelation in the 
response variable, by adopting a temporal block-permutation scheme. In both cases, 
the test is insignificant, see Appendix~\ref{app:permutation}. 

\begin{figure}[t]
\centering
\includegraphics[width=.71\linewidth]{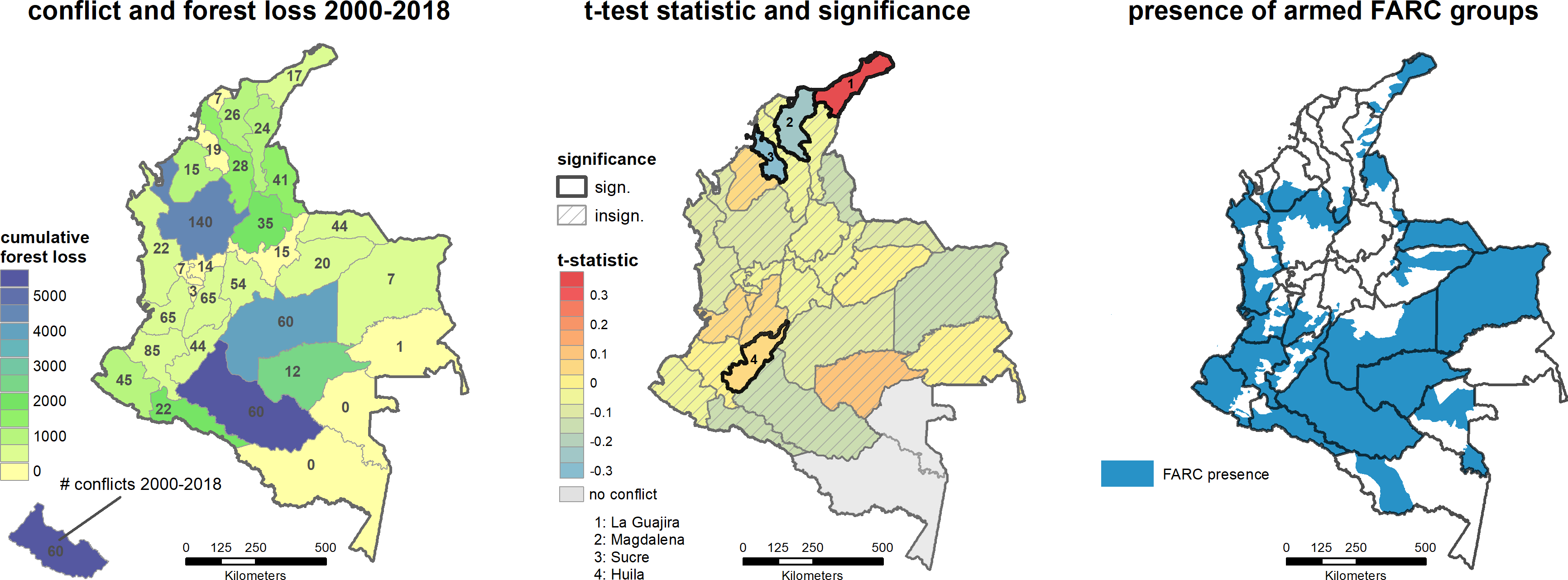}
\caption{
Regional analysis of conflict and forest loss in Colombia. The left panel shows the total forest loss and the total 
number of conflict events 2000--2018 at department level. The most severe incidences of forest 
loss occur in the northern Andean forests and on the northern borders of the Amazon region.
In the middle panel, we report for each department estimates $\hat{T}$ 
and test results for $H_0 : T = 0$, using the methodology described in Sections~\ref{sec:quant}~and~\ref{sec:testing}. 
We used a test level of $\alpha = 0.05$, and report significances without multiple-testing adjustment. In most departments, 
the estimated causal effect is negative (blue, conflict reduces forest loss), although mainly insignificant. We identify four departments 
with statistically significant results, hereof two with a positive causal effect (La Guarija and Huila) and two with a negative causal 
effect (Magdalena and Sucre). In total, there are 8 departments that are mostly controlled by FARC (above 75\% FARC presence, right panel). 
Out of these, 6 departments have a negative test statistic (conflict reduces forest loss).
}
\label{fig:regional}
\end{figure}

The above analysis provides an estimate for the average causal effect, see Equation~\eqref{eq:fAVE0}, 
which, in particular, averages over space. Given that Colombia is a country with 
high ambiental and socio-economic heterogeneity, 
we furthermore conduct an analysis at the department level (see Figure~\ref{fig:regional}). In fact, there is 
considerable spatial variation in the estimated causal effects, with significant positive as well as 
negative effects (Figure~\ref{fig:regional} middle).
This variation may be seen as evidence for an interaction effect between conflict and the assumed hidden confounders. 
In most departments, the estimated causal effect 
is negative (although mostly insignificant), meaning that conflict tends to decrease forest loss.
The strongest positive and significant causal influence is identified in the La Guajira department 
($\hat T = 0.398$, $P = 0.047$). 
Although this region is commonly associated with semi-arid to very dry conditions,
most conflicts occured in the South-Western areas, at the beginning of Caribbean tropical forests (see Figure~\ref{fig:col_fl}).
In fact, these zones have also been also identified by \cite{negret2019emerging} as having been 
strongly affected by deforestation pressure in the wake of conflict. Interestingly, the neighboring Magdalena department 
shows the opposite effect ($\hat{T} = -0.218$, $P = 0.004$).
The positive effect in the department of Huila ($\hat T = 0.095$, $P = 0.023$) is again in line with the findings by \cite{negret2019emerging} 
(based on a visual inspection of their attribution maps).
Out of the 8 departments that are mostly controlled by FARC (Figure~\ref{fig:regional} right), 
6 have a negative test statistic, meaning that conflict reduces forest loss. This can be explained in 
part by the internal governance of this group, where forest cover was a strategic advantage for both 
their own protection as well as for cocaine production.  Overall, of course, the peace-induced 
acceleration of forest loss has to be discussed with caution, and should not be interpreted 
reversely as if conflict per se is a measure of environmental protection 
\cite{clerici2020deforestation}.

\subsection{Interpretation in light of the Colombian peace process} \label{sec:conflict_peace}
\begin{figure}
\centering
\includegraphics[width=.48\linewidth]{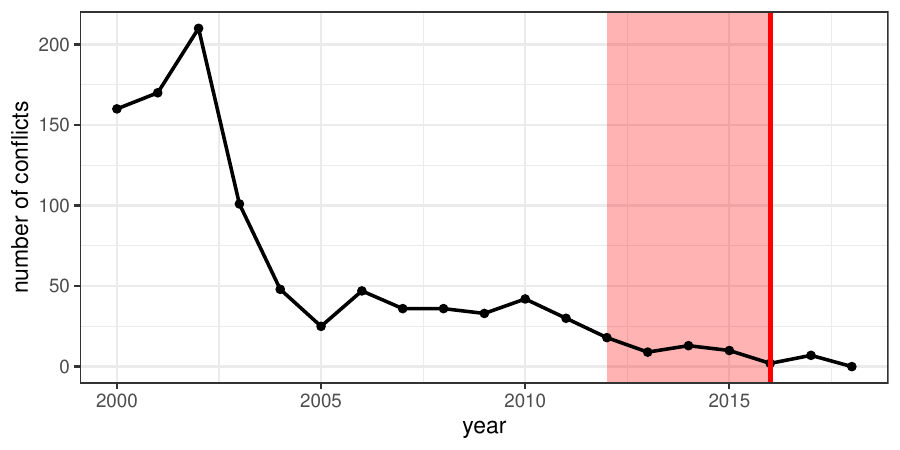}
$ $
\includegraphics[width=.48\linewidth]{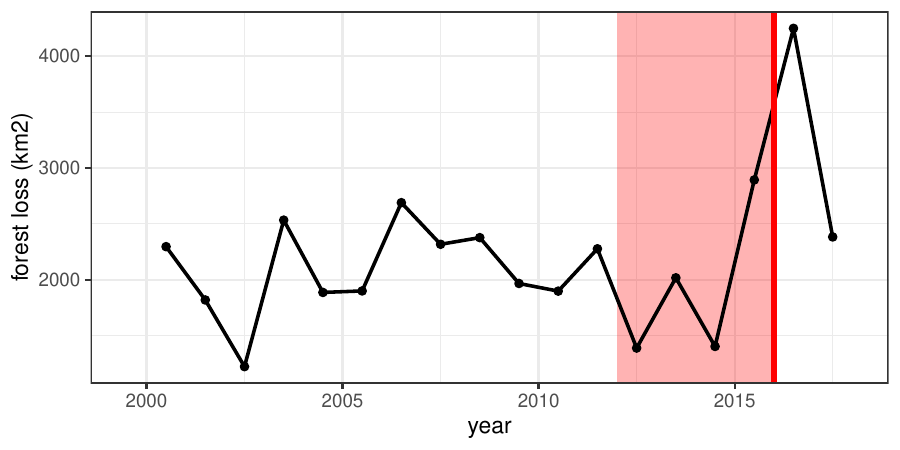}
\caption{Number of conflicts (left) and forest loss (right) in the years 2000--2018 in Colombia. 
In the right panel, the height of the curve at $x+0.5$ indicates the forest loss between years $x$ and $x+1$. 
Shaded areas mark the Colombian peace process (2012--2016).
 }
\label{fig:peace}
\end{figure}

In late 2012, negotiations that later would be known as the `Colombian peace process' started between the then president 
of Colombia and the strongest group of rebels, the FARC, lasting until 2016. 
Peace was declared by both parties upon a revised 
agreement in October 2016, and became effective in 2017.%
\footnote{The agreement was signed only by the FARC, while other guerilla groups remain active.}
The negotiations marked a steadily decreasing number of conflicts (Figure~\ref{fig:peace}, left).
Since this decrease %
is the result of governmental intervention, 
rather than a natural resolution of local tensions, 
the peace process provides an opportunity to verify the intervention effects estimated in Section~\ref{sec:conflict_results}. 
As can be seen in Figure~\ref{fig:peace} (right), Colombia experienced a steep increase in the total forest loss in the final phase of the peace negotiations. 
Although there may be several other factors which have contributed to this development, 
we observe that these results align with our previous finding of an overall negative causal effect of conflict on forest loss ($\hat T < 0$).

\section{Conclusions and future work}
\subsection{Methodology}
This paper 
introduces ways to discuss
causal inference for spatio-temporal data. From a methodological perspective, it contains three
main contributions: 
the definition of a class of causal models for multivariate spatio-temporal stochastic processes, 
a procedure for estimating causal effects within this model class, and
a non-parametric hypothesis test for the overall existence of such effects.
Our method allows for the influence of arbitrarily many latent confounders,
as long as these confounders do not vary across time (or space). 
We prove asymptotic consistency of our estimator, and verify this finding empirically using simulated data. 
Our results hold
under weak
assumptions on the data generating process, 
and do
not rely on any particular distributional properties of the data. 
We prove sufficient conditions under which these rather general assumptions hold true. 
The proposed testing procedure is based on data resampling and 
provably obtains valid level in finite samples. 

Our work can be extended into several directions.
We proved that Assumption~\ref{ass:LLN} holds for regularly sampled stationary Gaussian processes. Such settings
allow for the application of well-known theorems about stationary and ergodic sequences. We hypothesize 
that the assumption also holds in the more general case, where the marginalized process does not resemble a (collection of) 
stationary sequence(s), as long certain mixing conditions are satisfied. For example, we believe that if the original spatial
process $\B{H}^1$ is weakly stationary and mixing, 
Assumption~\ref{ass:LLN} holds under
any spatial sampling scheme $(s_n)_{n \in \N }$ with $\norm{s_n}_2 \to \infty$ as 
$n \to \infty$.

Our method assumes that all latent confounders are time-invariant. It would be interesting to see whether the error of our 
estimation procedure can be bounded when, violating the assumptions, there are unobserved, time-varying confounders, but 
such confounders vary only slowly over time. 
Likewise, it would be worthwile exploring how smoothness assumptions on the spatial variation of the hidden variables can 
be incorporated in the modeling process to gain statistical efficiency of our estimator. Such smoothness assumptions would also allow for alternative 
permutation schemes (data can be permuted spatially wherever $\B{H}$ is assumed to be constant) which could lead
to increased power of our hypothesis tests.

\subsection{Case study}
We have applied
our methodology to the problem of 
conflict and forest loss in Colombia. 
Conflicts 
are predictive of exceedances in forest loss, 
but we find no evidence of a 
causal relation when analyzing this problem on country level: 
once all (time-invariant) confounders are adjusted for, there is a negative but insignificant 
correlation between conflict and forest loss. 
Our analysis on department level suggests that this non-finding could be due to locally 
varying effects of opposite directionality, which 
would 
approximately
cancel out
in our final estimate. 
In most departments, we find negative (mostly insignificant) effects, 
although we also identify a few departments where conflict seems to 
increase forest loss. 
It would be interesting to further explore these differences in causal dependence, e.g., 
by adjusting for socio-political factors.
The overall negative influence of conflict on forest loss estimated 
by our method is in line with the observation that during the final phase of the peace process, which stopped 
many of the existing conflicts, the total forest loss in Colombia has increased. 
However, these results should be interpreted with caution. Overall, we find that, 
once all time-invariant confounders are adjusted for, conflicts have only weak 
explanatory power for predicting forest loss, and the potential causal effect 
is therefore likely to be small, compared to other drivers of forest loss. 
Further, our results rely on the assumption that all 
latent confounders are time-invariant, which, in practice, can only be expected to hold approximately. 
In Appendix~\ref{sec:timevar_conf_sim}, we explore how our estimator is affected by temporal 
variation in the latent confounders.

\section*{Acknowledgments}
This work contributes to the Global Land Programme, and was supported by 
Carlsberg Foundation (JP) and 
a research grant (18968) from 
VILLUM FONDEN (RC, JP).
MDM thanks ESA for funding the Earth System Data Lab. 
We are grateful to Steffen Lauritzen, Niels Richard Hansen and Niklas Pfister for valuable comments on the methodology, 
and to Lina Estupinan-Suarez and Alejandro Salazar for helpful discussions on the case study. We thank two anonymous referees 
and the associate editor for many helpful comments.

\appendix

\section{Existing causal methodology} \label{app:existing_causal_methods}
For independent and identically distributed (i.i.d.) and time series data,
that is, for data, for which the spatial information can be neglected,
causal models have been well-studied. 
Among the most widely used approaches are
structural causal models, 
causal graphical models, and the 
framework of potential outcomes \citep[see, e.g.,][]{bollen2014structural, pearl2009causality, 
peters2017elements,rubin1974estimating}.
Knowledge of the causal structure of a system
does not only provide us with 
cause-effect relationships; 
sometimes, it also allows us to 
quantify causal relations by
estimating
intervention effects from observational data. 
If, for example, we know 
that $W$ is causing $X$ and $Y$, and that $X$ is causing $Y$, 
procedures such as variable adjustment can be used to estimate the causal influence 
of $X$ on $Y$ 
from i.i.d.\ replications \citep{pearl2009causality, rubin1974estimating}.
While using a slightly different language, the same underlying causal deliberations
are the basis of many works
in 
econometrics, e.g., work related to 
generalized methods of moments and identifiability of parameters
\citep[e.g.,][]{hansen82, newey94}.
In this field, data are often assumed to have a time series structure \citep[e.g.,][]{hall05}.

For spatio-temporal data, several methods have been proposed
\citep[e.g.,][]{lozano2009spatial, luo2013spatio}.
These are 
mostly algorithmical approaches 
that extend
the concept of
Granger causality \citep{granger1980testing, Wiener1956}
to spatio-temporal data. They 
reduce the question of causality 
to predictability and a positive time lag.
In particular, these methods assume that there are no relevant unobserved variables (`confounders')
and they do not resolve the question of time-instantaneous causality between 
different points in space. 
More classical approaches for causal inference 
\citep[e.g.,][]{peirce1883theory, rosenbaum1983central, kelejian2004instrumental, delgado2015difference} require either 
a controlled study design, 
data from all confounding variables, 
a valid instrument, or 
a naturally occurring experiment, respectively, 
and are therefore not applicable to the case study considered in this paper.
Furthermore,
to the best of our knowledge,
existing work
does not provide 
a formal model for causality
for spatio-temporal 
data.
As a consequence, the precise definition 
of the target of inference,
the causal effect, 
remains vague.

\section{Extensions of our methodology} \label{sec:extensions}

\subsection{Observed time-varying confounders} \label{sec:timevar_conf}

For simplicity, we have until now 
assumed that the only confounders of $(X,Y)$ are the variables in $H$. 
Our method naturally extends to settings with observed (time- and space-varying) confounders. 
Let $W \in \R^p$ be some observed covariates, and consider a causal graphical model over $(\B{X}, \B{W}, \B{Y}, \B{H})$
with causal structure $[\B{Y} \given \B{X}, \B{W}, \B{H}] [\B{X} \given \B{W}, \B{H}] [\B{W}, \B{H}]$.
Similarly to Definition~\ref{defi:LSCM}, assume that $\B{W}$ and $\B{H}$ are weakly stationary, 
$\B{H}$ is time-invariant, and there exists a function $f : \R^{d+p+\ell+1} \to \R$ and an 
i.i.d.\ sequence $\boldsymbol{\epsilon}^1, \boldsymbol{\epsilon}^2, \dots$ of weakly stationary
error processes, independent of $(\B{X}, \B{W}, \B{H})$, such that
\begin{equation}
Y_s^t = f(X_s^t, W_s^t, H_s^t, \epsilon_s^t) \quad \text{ for all } (s,t) \in \R^2 \times \N.
\end{equation}
We define the average causal effect of $\B{X}$ on $\B{Y}$, for every $x \in \R^d$, by
\begin{equation*}
f_{\text{AVE}(X \to Y)}(x) = \E[f(x, W_0^1, H_0^1, \epsilon_0^1)].
\end{equation*}
It is straight-forward
to show that this function enjoys the same causal interpretation as given in Proposition~\ref{prop:causal_int}. 
Similarly to Proposition~\ref{prop:cov_adj}, $f_{\text{AVE}(X \to Y)}$ can be recovered from the full 
observational distribution over $(X_0^1, W_0^1, Y_0^1, H_0^1)$ by conditioning on $(X_0^1, W_0^1, H_0^1)$, 
followed by taking the expectation w.r.t.\ $(W_0^1, H_0^1)$. More precisely, we have\footnote{In this equation, we omit the sub- and superscripts for 
notational convenience. }
\begin{align*}
f_{\text{AVE}(X \to Y)}(x) 
&= \int_{\R^{p+\ell}} f_{Y \vert (X,W,H)}(x,w,h) \P_{(W,H)}(d(w,h)) \\
&= \int_{\R^\ell} \int_{\R^p} f_{Y \vert (X,W,H)}(x,w,h) \P_{W \vert (H = h)}(dw) \P_{H}(dh),
\end{align*}
where $f_{Y \vert (X,W,H)}$ is the regression function of $Y_s^t$ onto $(X_s^t, W_s^t, H^t_s)$. Motivated by the above equation, we estimate 
$f_{\text{AVE}(X \to Y)}$ by first estimating the regression function $f_{Y \vert (X,W,H)}$, and then approximating the two integrals. 
Since $\B{H}$ is time-invariant, we can, similarly to \eqref{eq:fAVEhat}, use the estimator
\begin{equation} \label{eq:fave_extension}
\hat f^{nm}_{\text{AVE}(X \rightarrow Y)}(\B{X}_n^m, \B{Y}_n^m, \B{W}_n^m)(x) := \frac{1}{n} \sum_{i = 1}^n \hat \E_{s_i} [\hat f^m_{Y \vert (X, W)}(\B{X}^m_{s_i}, \B{Y}^m_{s_i}, \B{W}_{s_i}^m)(x,W_0^1)],
\end{equation}
where $\hat f^m_{Y \vert (X, W)}(\B{X}^m_{s_i}, \B{Y}^m_{s_i}, \B{W}_{s_i}^m)$ is an estimate of $(x,w) \mapsto f_{Y \vert (X,W,H)}(x,w,h_{s_i})$
computed from the observed data at location $s_i$, and $\hat \E_{s_i}$ is the expectation w.r.t.\ the empirical distribution of $\B{W}_{s_i}^m$
(which serves as an approximation of the inner integral).

\subsection{Temporally lagged causal effects} \label{sec:temp_lagged}
We can incorporate temporally lagged causal effects by allowing the function 
$f$ in \eqref{eq:f} to depend on past values of the predictors. That is, we model 
a joint causal influence of the past $k \geq 1$ temporal instances of the predictors by
assuming the existence of a function $f : \R^{d \cdot k + \ell + 1} \to \R$ and 
an i.i.d.\ sequence $\boldsymbol{\epsilon}^1, \boldsymbol{\epsilon}^2, \dots $
such that 
\begin{equation*}
Y_s^t = f(X_s^{t-k+1}, \dots, X_s^t, H_s^t, \epsilon_s^t) \quad \text{ for all } s \in \R^2 \text{ and } t \geq k.
\end{equation*}
In this case, the average causal effect \eqref{eq:fAVE0} is a function $\R^{d \cdot k} \to \R$
which can be estimated, similarly to \eqref{eq:fAVEhat}, using a regression estimator 
$\hat{f}^m_{Y \vert X}$ of $Y_s^t$ onto $(X_s^{t-k+1}, \dots, X_s^t)$.

\subsection{Exchanging the role of space and time} \label{sec:exchange_st}
We have assumed that the hidden confounders do not vary across time. This assumption
allowed us to estimate the regression $x \mapsto \E[Y_s^t \given X_s^t = x, H_s^t = h]$
at all unobserved values $h$. In fact, our method can be formulated in more general terms. 
If $(\B{X}, \B{Y}, \B{H})$ is a multivariate process defined on some general, possibly random,
index set $\mathcal{I} = I_1 \times \cdots \times I_p$ (see the definition of a data cube \citep{mahecha2020earth}), 
it is enough to require $\B{H}$ to be
invariant across one (or several) of the dimensions in $\mathcal{I}$. Similarly to \eqref{eq:fAVEhat}, 
the idea is then to estimate the dependence of $\B{Y}$ on $(\B{X}, \B{H})$ along these invariant 
dimensions, followed by an aggregation across the remaining dimensions. In case of a spatio-temporal
process, for example, our method also applies if the hidden variables are constant across space, 
rather than time.

\section{Alternative definition of the average causal effect} \label{app:steffen}

In our definition of the average causal effect \eqref{eq:fAVE0}, we take the expectation 
with respect to the hidden variables $H$. By the assumption of time-invariance, 
however, there is only a single replication of the spatial process $\B{H}^1$. One may 
argue that it is more relevant to define the inferential target in terms of that one
realization, rather than in terms of a distribution over possible alternative outcomes
which will never manifest themselves. This leads to the alternative definition of the 
average causal effect %
\begin{equation*}
x \mapsto
\lim_{S \to \infty} \frac{1}{(2S)^2} \int_{[-S,S]^2} \E[f(x,h^1_s, \epsilon_0^1)] \, ds = \lim_{S \to \infty} \frac{1}{(2S)^2} \int_{[-S,S]^2} f_{Y \vert (X,H)}(x,h^1_s) \, ds,
\end{equation*}
assuming that the above limits exist. Under the assumption of ergodicity of $\B{H}^1$, 
the above expression coincides with Definition~\ref{defi:fave}, but it is learnable from 
data, via the estimator introduced in Section~\ref{sec:estimator}, even if this is not the case. 
Here, we choose the formulation in Definition~\ref{defi:fave}
because we found that it results in a more comprehensible theory.

\section{Sufficient conditions for Assumptions~\ref{ass:LLN}~and~\ref{ass:fhat}} \label{sec:suff_cond}
For Assumption~\ref{ass:LLN}, we consider a stationary Gaussian setup.
By considering a regular spatial sampling scheme, we can make use of 
standard ergodic theorems for stationary and ergodic time series. 
\begin{prop}[Sufficient conditions for Assumption~\ref{ass:LLN}] \label{prop:LLN}
Assume that $\B{H}^1$ is a stationary multivariate Gaussian process with covariance function 
$C :\R^2 \to \R^{\ell \times \ell}$, i.e., $C(h) = \emph{Cov}(H^1_s, H^1_{s+h})$ for all 
$s, h \in \R^2$. Assume that $C(h) \to 0$ entrywise as $\norm{h}_2 \to \infty$. 
Consider a regular grid $\{s^1_1, s^1_2, \dots \} \times \{s^2_1, s^2_2, \dots, s^2_m \} \subset \R^2$, 
where $s^1_1 < s^1_2 < \cdots$ are equally spaced, and let $(s_n)_{n \in \N}$ be the spatial sampling scheme 
for every $i \in \N$ and $j \in \{1, \dots, m\}$ given as $s_{(i-1)m+j} = (s^1_i, s^2_j)$. 
Then, the process $(H^1_{s_n})_{n \in \N}$ satisfies Assumption~\ref{ass:LLN}.
\end{prop}
The proof of the above proposition uses mixing conditions on the sequence $(H^1_{s_n})_{n \in \N}$, 
which rely on the fact that $\norm{s_n}_2 \to \infty$. 
In spatial statistics, these types of asymptotic regimes are known as `increasing domain asymptotics'. 
To ensure Assumption~\ref{ass:LLN} in a setting where $(s_n)_{n \in \N}$ corresponds to an increasingly 
fine sampling of some bounded domain (`in-fill asymptotics'), we believe that a different set of assumptions 
on the process $\B{H}^1$ would be necessary.

For Assumption~\ref{ass:fhat}, we consider the slightly stronger version formulated conditionally on $\B{H}$. 
We let $\mathcal{H} \subset \mathcal{Z}_\ell$ 
denote the set of all functions $\B{h} : \R^2 \times \N \to \R^\ell$ that are constant in the
time-argument. Since $\B{H}$ is time-invariant, we have that $\P(\B{H} \in \mathcal{H}) = 1$, 
and it therefore suffices to prove the statement for all $\B{h} \in \mathcal{H}$. 
Below, we use, for every $\B{h} \in \mathcal{H}$, $\P_\B{h}$ to denote the conditional distribution $\P(\cdot \given \B{H} = \B{h})$
and $\E_\B{h}$ for the expectation with respect to $\P_\B{h}$.

We now make some structural
assumptions on the function $f$ in \eqref{eq:f}, which allow us to parametrically estimate the regressions 
$x \mapsto f_{Y \vert (X,H)}(x,h)$.
Let $\{ \varphi_1, \dots, \varphi_p \}$ be a known basis of continuous functions on $\R^d$, 
and with $\varphi_1 \equiv 1$ an intercept term. 
With $\varphi := (\varphi_1, \dots, \varphi_p)$, we make the 
following assumptions on the underlying LSCM.
\begin{itemize}
	\item[(L1)] There exist functions $f_1 : \R^\ell \to \R^{p}$ and 
	$f_2 : \R^{\ell+1} \to \R$ such that Equation~\eqref{eq:f} 
	splits into 
	\begin{equation*}
	Y_s^t = \varphi(X_s^t)^\top f_1(H_s^t) + f_2(H_s^t, \epsilon_s^t) \qquad \text{ for all } (s,t) \in \R^2 \times \N,
	\end{equation*}
	and such that for all $h \in \R^\ell$, $f_2(h, \epsilon_0^1)$ has finite second moment.
\end{itemize}
For every $s,t$, define $\xi_s^t = f_2(H_s^t, \epsilon_s^t)$. We can w.l.o.g.\ assume that for all $s$, $t$ and $h$, $\E[\xi_s^t \given H_s^t = h] = 0$. 
(Since $\varphi_1 \equiv 1$, this can always be accommodated by adding $h \mapsto \E[\xi_s^t \given H_s^t = h]$
to the first coordinate of $f_1$.) 
For every fixed $\B{h} \in \mathcal{H}$ and $s \in \R^2$, assumption (L1) 
says that, under 
$\P_\B{h}$, $(\B{X}_s, \B{Y}_s)$ follows a simple regression model, where $\E[Y_s^t \given X_s^t]$
depends linearly on $\varphi(X_s^t)$. 
For arbitrary but fixed $h_s^1$, 
we can therefore estimate $x \mapsto \E[Y_s^1 \given X_s^1 = x, H_s^1 = h_s^1]$
using standard OLS estimation. For every $s \in \R^2$ and $m \in \N$, let $\B{\Phi}_s^m \in \R^{m \times p}$ be the design matrix 
given by $(\B{\Phi}_s^m)_{ij} = \varphi_j(X_s^i)$. 
We define an estimator $\hat{f}_{Y \vert X} = (\hat{f}^m_{Y \vert X})_{m \in \N}$, for every $x \in \R^d$ and $m \in \N$ by 
\begin{equation} \label{eq:fgammahat}
\hat f^m_{Y \vert X}(\B{X}_s^m, \B{Y}_s^m)(x) = \varphi(x)^\top \hat{\gamma}_s^m,
\end{equation}
where 
$
\hat{\gamma}_s^m := ((\B{\Phi}_s^m)^\top \B{\Phi}_s^m)^{-1} (\B{\Phi}_s^m)^\top \B{Y}_s^m.
$
To formally prove consistency of \eqref{eq:fgammahat}, we need some regularity conditions on the predictors $\B{X}$. 
\begin{itemize}
	\item[(L2)] For all $\B{h} \in \mathcal{H}$, $s \in \R^2$ and $\delta > 0$, it holds that 
	\begin{equation*}
	\lim_{m \to \infty} \P_\B{h}(\norm{\tfrac{1}{m} (\B{\Phi}_s^m)^\top \boldsymbol{\xi}_s^m}_2 > \delta) = 0. 
	\end{equation*}	
	\item[(L3)] For all $\B{h} \in \mathcal{H}$, $s \in \R^2$, there exists $c > 0$ such that 
	\begin{equation*} \label{eq:lambdamin}
	\lim_{m \to \infty} \P_\B{h}(\lambda_{\text{min}} \left( \tfrac{1}{m} (\B{\Phi}_s^m)^\top \B{\Phi}_s^m \right) \leq c ) = 0,
	\end{equation*}
	where $\lambda_\text{min}$ denotes the minimal eigenvalue. 
\end{itemize}
We first state the result and discuss assumptions (L2) and (L3) afterwards.
\begin{prop}[Sufficient conditions for Assumption~\ref{ass:fhat}] \label{prop:fhat}
Assume that $(\B{X}, \B{Y}, \B{H})$ come from an LSCM satisfying (L1)--(L3). 
Then, Assumption~\ref{ass:fhat} is satisfied with $\mathcal{X} = \R^d$ and with $\hat{f}^m_{Y \vert X}$ 
as defined in \eqref{eq:fgammahat}. 
\end{prop}
Since for every $(s,t) \in \R^2 \times \N$ and $\B{h} \in \mathcal{H}$, $X_s^t$ and $\xi_s^t$ are independent under $\P_\B{h}$
with $\E_\B{h}[\xi_s^t] = 0$, 
(L2) states a natural LLN-type condition, which is satisfied under suitable constraints on the temporal dependence structure in $\B{X}$, and on its variance.
Assumption~(L3) says that, with probability tending to one, the matrix $\tfrac{1}{m} (\B{\Phi}_s^m)^\top \B{\Phi}_s^m$ is bounded away from singularity as $m \to \infty$. 
This is in particular satisfied if $\tfrac{1}{m} (\B{\Phi}_s^m)^\top \B{\Phi}_s^m$ converges in probability entrywise to some matrix which is strictly positive definite. 
In Appendix~\ref{app:examples}, we give two examples in which this is the case.

\section{Examples satisfying conditions L1 and L2} \label{app:examples}

Let $(\B{X}, \B{Y}, \B{H})$ come from an LSCM satisfying condition (L1) described in Appendix~\ref{sec:suff_cond}. 
Below, we give two examples of distributions over $(\B{X}, \B{H})$ for which also conditions (L2)~and~(L3) hold true. 
In both cases, 
$\tfrac{1}{m} (\B{\Phi}_s^m)^\top \B{\Phi}_s^m$ converges in probability to some limit matrix of the form $\E_\nu[\varphi(X) \varphi(X)^\top]$
for some measure $\nu$ with full support on $\R^d$. To see that $\E_\nu[\varphi(X) \varphi(X)^\top]$ is strictly positive definite, 
let $v \in \R^p$ be such that $0 = v^\top \E_\nu[\varphi(X) \varphi(X)^\top] v = \E_\nu[ \norm{\varphi(X)^\top v}_2^2]$. By continuity of 
$\varphi$, it follows that $\varphi^\top v \equiv 0$, and the linear independence of $\varphi_1, \dots, \varphi_p$ implies that $v = 0$.

\begin{ex}[Temporally ergodic $\B{X}$] \label{ex:ergodic}
Let $(\B{X}, \B{Y}, \B{H})$ come from an LSCM satisfying Assumption~(L1). 
Assume that for every $\B{h} \in \mathcal{H}$ and $s \in \R^2$, it holds that under $\P_\B{h}$, 
$\B{X}_s$ is a stationary and mixing process with a marginal distribution that has full support on $\R^d$
(e.g., a vector autoregessive process with additive Gaussian noise). Assume furthermore that $\E_\B{h}[\card{\xi_s^1}^2] < \infty$ 
and $\E_\B{h}[\card{\varphi_i(X_s^1)}^2 < \infty]$ 
for all $i \in \{1, \dots, p\}$. Analogously to the proof of Proposition~\ref{prop:LLN}, we can then
show that for each $i,j \in \{1, \dots, p\}$, the sequences $(\varphi_i(X_s^t) \xi_s^t)_{t \in \N}$ and $(\varphi_i(X_s^t) \varphi_j(X_s^t))_{t \in \N}$
are ergodic under $\P_\B{h}$, and it follows that 
\begin{equation*}
(\tfrac{1}{m} (\B{\Phi}_s^m)^\top \boldsymbol{\xi}_s^m)_{i} = \frac{1}{m} \sum_{t=1}^m \varphi_i(X_s^t) \xi_s^t \to \E_\B{h}[ \varphi_i(X_s^1) \xi_s^1]
\end{equation*}
and 
\begin{equation*}
(\tfrac{1}{m} (\B{\Phi}_s^m)^\top \B{\Phi}_s^m)_{ij} = \frac{1}{m} \sum_{t=1}^m \varphi_i(X_s^t) \varphi_j(X_s^t) \to \E_\B{h}[ \varphi_i(X_s^1) \varphi_j(X_s^1)]
\end{equation*}
as $m \to \infty$ in probability under $\P_\B{h}$. Since for all $s \in \R^2$, $\E_\B{h}[ \varphi(X_s^1) \xi_s^1] = \E_\B{h}[ \varphi(X_s^1)] \cdot \E_\B{h}[\xi_s^1] = 0$
and $\E_\B{h}[\varphi(X_s^1) \varphi(X_s^1)^\top] \succ 0$, the above implies (L2) and (L3).
\end{ex}

\begin{ex}[Temporally independent $\B{X}$ with convergent mixture distributions] \label{ex:mixing}
Let $(\B{X}, \B{Y}, \B{H})$ come from an LSCM satisfying Assumption~(L1) for some bounded functions $\varphi_1, \dots, \varphi_p$.
Assume that for every $s \in \R^2$, the variables $X_s^1, X_s^2, \dots $ are conditionally
independent given $\B{H}$ (they are not required to be identically distributed), and that 
for every $\B{h} \in \mathcal{H}$, the sequence of mixture distributions 
\begin{equation} \label{eq:mixture}
\P_{s, \B{h}}^{m} := \frac{1}{m} \sum_{t=1}^m \P_{X_s^t \vert \B{H} = \B{h}}, \quad m \in \N,
\end{equation}
converges, for $m \rightarrow \infty$, 
weakly towards some limit measure $\P_{s, \B{h}}^{\infty}$ with full support on $\R^d$. 
Then, conditions (L1)~and~(L2) are satisfied. 
To see this, let $\B{h} \in \mathcal{H}$ and $s \in \R^2$ be fixed for the rest of this example. 
Let $m \in \N$ and $\delta > 0$. Since $\E_\B{h}[(\B{\Phi}_s^m)^\top \boldsymbol{\xi}_s^m] = 0$, it follows from Chebychev's inequality 
that for all $i \in \{1, \dots, p\}$,
\begin{equation*}
\P_\B{h}(\card{\tfrac{1}{m} ((\B{\Phi}_s^m)^\top \boldsymbol{\xi}_s^m)_i} > \delta) \leq \frac{1}{\delta^2} \emph{Var}_\B{h} ( \frac{1}{m}((\B{\Phi}_s^m)^\top \boldsymbol{\xi}_s^m)_i) = \frac{\E_\B{h}((\xi_0^1)^2)}{\delta^2 m} \underbrace{\E_{s, \B{h}}^{m}[\varphi_i(X)^2]}_{\emph{unif. bounded}} \to 0,
\end{equation*}
as $m \to \infty$, showing that (L2) is satisfied.
To prove (L3), let $M^{m} := \E_{s, \B{h}}^{m}[\varphi(X) \varphi(X)^\top]$, $m \in \N$, and $M^\infty := \E_{s, \B{h}}^{\infty}[\varphi(X) \varphi(X)^\top]$ (to simplify notation, we here omit the implicit dependence on $\B{h}$ and $s$).
By assumption on $(\P_{s, \B{h}}^{m})_{m \in \N}$, $M^m$ converges entrywise to $M^\infty$ as $m \to \infty$. Together with another application of Chebychev's inequality, it follows that for all $i,j \in \{1, \dots, p\}$, 
\begin{align*}
\P_\B{h}(\card{\tfrac{1}{m} ((\B{\Phi}_s^m)^\top \B{\Phi}_s^m)_{ij} - M^\infty_{ij}} > 2\delta) 	&\leq \P_\B{h}(\card{\tfrac{1}{m} ((\B{\Phi}_s^m)^\top \B{\Phi}_s^m)_{ij} - M^{m}_{ij}} > \delta) + \P_\B{h}(\card{M^{m}_{ij} - M^\infty_{ij}} > \delta) \\
																																					&\leq \frac{1}{\delta^2m} \underbrace{\E_{s, \B{h}}^{m}[\varphi_i(X)^2 \varphi_j(X)^2]}_{\emph{unif. bounded}} + \underbrace{\P_\B{h}(\card{M^{m}_{ij} - M^\infty_{ij}} > \delta)}_{=0 \emph{ for } m \emph{ large}} \\
																																					&\to 0 \quad \text{ as } m \to \infty,
\end{align*}
showing that $\tfrac{1}{m} ((\B{\Phi}_s^m)^\top \B{\Phi}_s^m)$ converges entrywise to $M^\infty \succ 0$ in probability under $\P_\B{h}$, and (L3) follows. 
\end{ex}

\begin{figure}
\centering
\includegraphics[width=\linewidth]{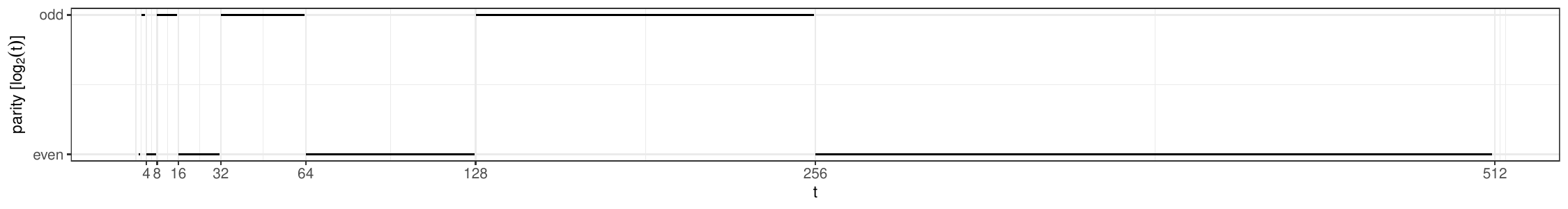}
\caption{
Visualization of the example in Remark~\ref{rem:mixing}. Whenever the parity of $\lfloor \log_2 t \rfloor$
changes from even to odd, the entire mass of $\P_{X_s^t \vert \B{H} = \B{h}}$ moves from $(-\infty, -1]^d$ to $[1, \infty)^d$, and vice versa. 
In this case, the mixture $\P_{s, \B{h}}^m$ in \eqref{eq:mixture} does not converge, and the consistency in Proposition~\ref{prop:fhat} does not hold in general.
}
\label{fig:remark}
\end{figure}

\begin{rem}[Necessity of the convergence of mixtures] \label{rem:mixing}
The convergence assumption on $\P_{s, \B{h}}^{m}$ is crucial for obtaining the above consistency result. 
It is easy to construct examples of $(\P_{X^t_s \vert \B{H} = \B{h}})_{t \in \N}$ where this assumption
fails to hold. For example, let 
$\P_\B{h}(X_s^t \in (-\infty, -1]^d) = 1$ whenever $\lfloor \log_2 t \rfloor$ is even, 
and $\P_\B{h}(X_s^t \in [1, \infty)^d) = 1$ whenever $\lfloor \log_2 t \rfloor$ is odd. 
This construction is visualized in Figure~\ref{fig:remark}.
Then, both sequences $(\P_\B{h}(X_s^t \in (-\infty, -1]^d))_{t \in \N}$ and $(\P_\B{h}(X_s^t \in [1, \infty)^d))_{t \in \N}$
alternate between zero and one, with a frequency chosen such that for all $k \geq 2$ even, 
$\P_{s, \B{h}}^{2^{k}-1}([1, \infty)^d) = 2/3$, and for all $k \geq 3$ odd, $\P_{s, \B{h}}^{2^{k}-2}((- \infty, -1]^d) = 2/3$, 
showing that $\P_{s, \B{h}}^{m}$ does not converge.
In this case, the dataset $\{(X_s^t, Y_s^t) \st t \in \{1, \dots, m\} \}$ alternates between mostly containing 
pairs $(X_s^t, Y_s^t)$ with $X_s^t \in (-\infty, -1]^d$ and mostly containing pairs $(X_s^t, Y_s^t)$ with $X_s^t \in [1, \infty)^d$.
If the functional dependence of $Y_s^t$ on $X_s^t$ differs between these two domains, the estimator $\hat{f}^m_{Y \vert X}$ does therefore
not converge in general.
\end{rem}

\begin{rem}[Conditions implying the convergence of mixtures] \label{rem:mixing2}
We can make the convergence assumption on $\P_{s, \B{h}}^{m}$ more
concrete in the case where the distributions in $(\P_{X^t_s \vert \B{H} = \B{h}})_{t \in \N}$ differ only in their respective
mean vectors. Assume there exist functions $\mu_s^t : \mathcal{Z}_\ell \to \R^d$ and $g_s: \mathcal{Z}_\ell \times \R^d \to \R^d$,
$(s,t) \in \R^2 \times \N$, and a $d$-dimensional error process $\boldsymbol{\zeta} \indep \B{H}$, 
such that for each $s \in \R^2$, $\zeta_s^1, \zeta_s^2, \dots $ are i.i.d., and such that for all $(s,t) \in \R^2 \times \N$
it holds that $X_s^t = \mu_s^t(\B{H}) + g_s(\B{H}, \zeta_s^t)$. Assume furthermore that for each $\B{h} \in \mathcal{H}$ and $s \in \R^2$, 
$g_s(\B{h}, \zeta_s^0)$ has strictly positive density $f_{s, \B{h}}$ w.r.t.\ the Lebesgue measure on $\R^d$. We can then ensure convergence 
of the mixture distributions $\P_{s, \B{h}}^{m}$ by requiring that for each $\B{h} \in \mathcal{H}$ and $s \in \R^2$ there exists 
some density function $f^{\text{mix}}_{s, \B{h}}$ on $\R^d$, such that for all $x \in \R^d$ it holds that\footnote{By slight abuse of notation, we use 
$(-\infty, x]$ to 
denote the product set $\varprod_{i=1}^d (-\infty, x_i]$.}  
\begin{equation*}
\lim_{m \to \infty} \frac{1}{m} \sum_{t=1}^m \mathbbm{1}_{(-\infty, x]}(\mu_s^t(\B{h})) = \int_{(-\infty, x]} f^{\text{mix}}_{s, \B{h}}(z) dz.
\end{equation*}
(Intuitively, this equation states that, in the limit $m \to \infty$, the set $\{\mu_s^t(\B{h}) \st t \in \{1, \dots, m\}\}$
looks like an i.i.d.\ sample drawn from the distribution with density $f^{\text{mix}}_{s, \B{h}}$.)
For all $\B{h} \in \mathcal{H}$, $s \in \R^2$, $m \in \N$ and $x \in \R^d$, we then have 
\begin{align*}
\P_{s, \B{h}}^m((- \infty, x]) 	&= \frac{1}{m} \sum_{t=1}^m \int_{(- \infty, x]} f_{s, \B{h}}(v- \mu_s^t(\B{h})) dv \\
												&= \int_{\R^d} \frac{1}{m} \sum_{t=1}^m f_{s, \B{h}}(v- \mu_s^t(\B{h})) \mathbbm{1}_{(- \infty, x]}(v) dv \\
												&= \int_{\R^d} f_{s, \B{h}}(v) \frac{1}{m} \sum_{t=1}^m  \mathbbm{1}_{(- \infty, x-v]}(\mu_s^t(\B{h})) dv,									
\end{align*}
and it follows from the dominated convergence theorem that 
\begin{align*}
\lim_{m \to \infty} \P_{s, \B{h}}^m((- \infty, x]) 	&= \int_{\R^k} f_{s, \B{h}}(v) \int_{(- \infty, x-v]} f^{\text{mix}}_{s, \B{h}}(z) dz dv \\
																			&= \int_{(- \infty, x]} \int_{\R^k}  f_{s, \B{h}}(v)  f^{\text{mix}}_{s, \B{h}}(z-v) dv dz \\
																			&= \int_{(- \infty, x]} (f_{s, \B{h}} \ast f^{\text{mix}}_{s, \B{h}})(z) dz,
\end{align*}
showing that $\P_{s, \B{h}}^m$ converges weakly to the measure with 
the convoluted
density $f_{s, \B{h}} \ast f^{\text{mix}}_{s, \B{h}}$. 
Since $f_{s, \B{h}}$ is strictly positive, this measure has full support on $\R^d$.
\end{rem}

\section{Simulation experiments}

\subsection{Observed and unobserved time-varying confounders (synthetic data)} \label{sec:timevar_conf_sim}
\begin{figure}
\centering
\includegraphics[width=\linewidth]{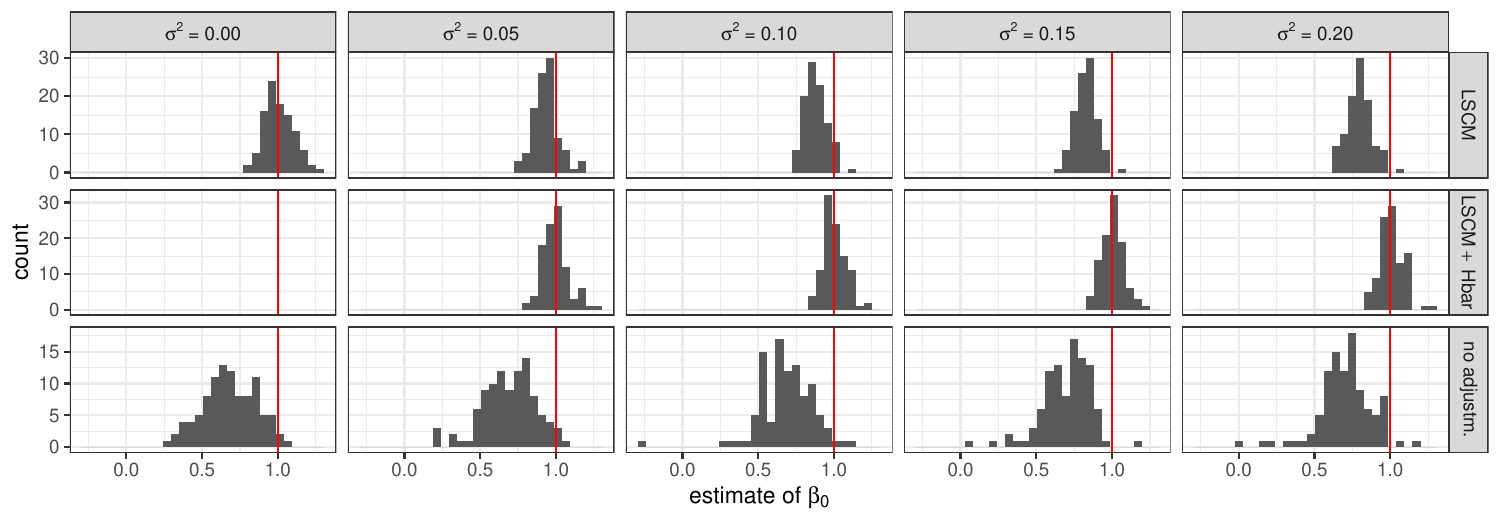}

\includegraphics[width=\linewidth]{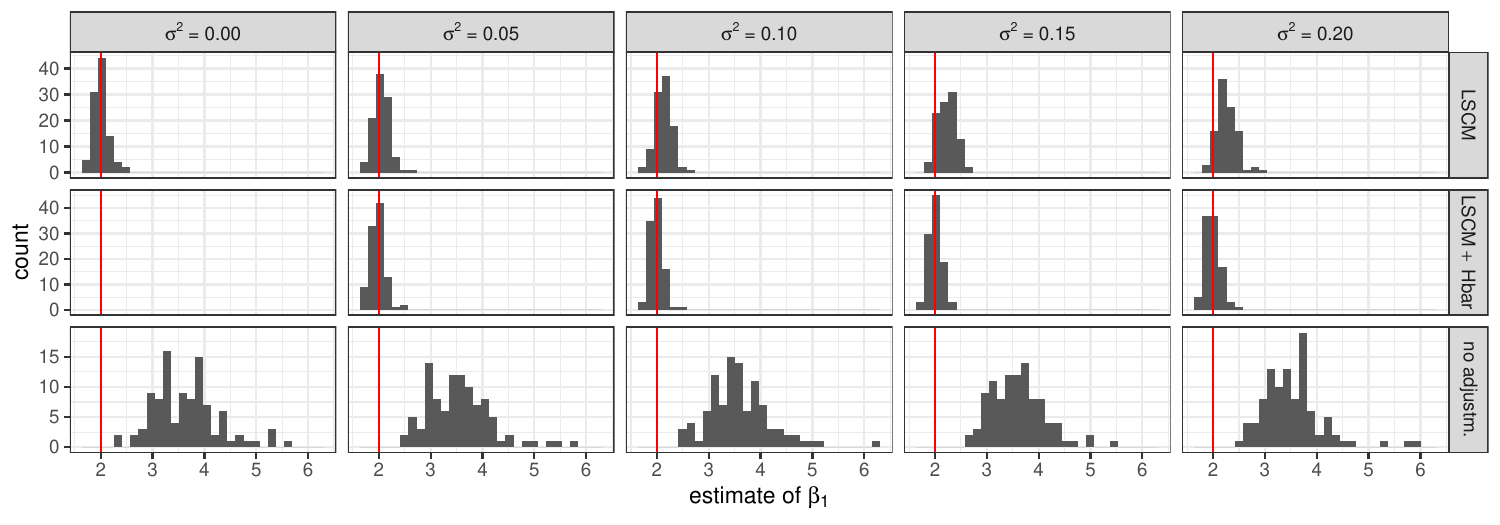}
\caption{Estimates of $\beta_0$ (top) and $\beta_1$ (bottom) for the simulation experiment described in Appendix~\ref{sec:timevar_conf_sim}. 
The true values ($\beta_0 = 1$ and $\beta_1 = 2$) are indicated by vertical red lines. If all confounders are time-invariant ($\sigma^2 = 0$), 
our procedure (`LSCM') correctly estimates the causal parameters. As the temporal variation of the confounder $\bar{H}_s^t$ increases (that is, as 
$\sigma^2$ becomes larger), our procedure becomes biased. If $\bar{H}_s^t$ is observed, it can readily be integrated into our method, 
see Appendix~\ref{sec:timevar_conf}. In this example, the resulting procedure (`LSCM + Hbar') yields empirically unbiased estimators. For $\sigma^2 = 0$, 
this method is undefined, since the location-wise regression of $Y_s^t$ onto $(X_s^t, H_s^t)$ is not identified.}
\label{fig:timevar_conf}
\end{figure}

If some of the confounders of $(X_s^t, Y_s^t)$ are not constant across time, the original version of our 
estimator defined in \eqref{eq:fAVEhat} cannot be expected to be consistent. Given observations from the 
time-varying confounders, our method can be extended using the approach described in Appendix~\ref{sec:timevar_conf}. 
We now investigate empirically 
how the extension can be used, 
and what happens if the time-varying confounders are ignored. We use a setup 
similar to Example~\ref{ex:sim}, but where we allow the confounder $\bar{\B{H}}$ to vary across time.
Let $\boldsymbol{\zeta}, \boldsymbol{\psi}, \boldsymbol{\xi}^{t}, \boldsymbol{\epsilon}^{t}$, 
$t \in \N$, be independent versions
of a univariate stationary spatial
Gaussian process with mean $0$ and covariance function $u \mapsto  e^{-\tfrac{1}{2} \norm{u}_2}$. 
In addition, let $\boldsymbol{\chi}$ be a spatio-temporal process consisting of i.i.d.\ standard Gaussian random variables.
Similarly to Example~\ref{ex:sim}, we define a marginal distribution over $\B{H} = (\bar{\B{H}}, \tilde{\B{H}})$ and 
conditional distributions $\B{X} \given \B{H}$ and 
$\B{Y} \given (\B{X}, \B{H})$ by specifying that for all $(s,t) \in \R^2 \times \N$,
\begin{align*}
H_s^t &= (\bar{H}_s^t, \tilde{H}_s^t) = (\sigma  \cdot \chi_s^t + a(\sigma ) \cdot \zeta_s, 1 + b(\sigma ) \cdot \zeta_s + c(\sigma ) \cdot \psi_s), \\
X_s^t &= \exp(-\norm{s}_2^2/1000) + (0.2 + 0.1 \cdot \sin( 2 \pi t / 100)) \cdot \bar{H}_s^t \cdot \tilde{H}_s^t + 0.5 \cdot \xi_s^t, \\
Y_s^t &= (1.5 + \bar{H}_s^t \cdot \tilde{H}_s^t) \cdot X_s^t+ (\bar{H}_s^t)^2 + \card{\tilde{H}_s^t} \cdot \epsilon_s^t.
\end{align*}
Here, $\sigma  \in [0,1)$ controls the temporal variation of $\bar{\B{H}}$, and the functions $a(\sigma ) := \sqrt{1-\sigma^2}$, 
$b(\sigma ) := \frac{1}{2 a(\sigma )}$ and $c(\sigma ) := \sqrt{1-b(\sigma )^2}$ are chosen such that for all $\sigma , s, t$, $\bar{H}_s^t$
and $\tilde{H}_s^t$ are standard Gaussian with $\E[\bar{H}_s^t \cdot \tilde{H}_s^t] = \frac{1}{2}$. For all $\sigma $, the average 
causal effect is the linear function $x \mapsto \beta_0 + \beta_1 x$, 
with $\beta_0 := \E[(\bar{H}_0^1)^2] = 1$ and $\beta_1 := 1.5 + \E[\bar{H}_0^1 \cdot \tilde{H}_0^1] = 2$. We consider the same spatial sampling 
scheme as in Example~\ref{ex:sim}, and generate data sets for a fixed sample size of $m=50$ and $n=1500$. 
For every data set, 
we compute an estimate of $\beta_0$ and $\beta_1$ using an estimator of the form \eqref{eq:fave_extension}, which uses data from $\bar{\B{H}}$ to adjust for 
both observed time-varying confounding and unobserved time-invariant confounding using the 
procedure described in Appendix~\ref{sec:timevar_conf}. (For the step where $Y_s^t$ is regressed onto $(X_s^t, \bar{H}_s^t)$, we assume 
a linear regression model which includes 
only the additive terms $X_s^t$, $X_s^t \cdot \bar{H}_s^t$ and $(\bar{H}_s^t)^2$, and no intercept.) 
For comparison, we also show results for the original estimator \eqref{eq:beta0hat} which assumes that all confounders are time-invariant, 
and from a classical linear regression which 
ignores the existence of confounders altogether. The results are shown in Figure~\ref{fig:timevar_conf}. If $\sigma^2 = 0$, 
that is, all confounders are time-invariant, our procedure (`LSCM') correctly estimates $\beta_0$ and $\beta_1$. As the variance $\sigma^2$ of the temporally
varying term of $H_s^t$ increases, our procedure becomes slightly biased, but still gives better results than ignoring the confounders (`no adjustm.'). 
The extension of our method (`LSCM + Hbar'), which assumes that $\bar{H}_s^t$ is observed, is empirically unbiased in all scenarios.

\subsection{Generating semi-real data} \label{sec:semireal}
We conduct two semi-real experiments to assess the performance of our methodology for the specific data application introduced in Section~\ref{sec:conflict_cont}.
We use the original data for conflict $\B{X}$, and generate new responses $\tilde{\B{Y}}$ using
the following approach. For different values of $\beta \in \R$, we generate, for every $s,t$, 
a new response $\tilde{Y}_s^t$ from the equation
\begin{equation} \label{eq:genY}
\tilde{Y}_s^t := \beta X_s^t + \tilde{\epsilon}_s^t.
\end{equation}
Here, the noise variable $\tilde{\epsilon}_s^t$ is drawn from the empirical forest loss distribution at location $s$. More precisely, the variables
in $\tilde{\boldsymbol{\epsilon}} = (\tilde{\epsilon}_s^t)_{(s,t) \in \{s_1, \dots, s_n\} \times \{1, \dots, m\}}$ are conditionally independent given 
$\B{Y}$ and have (conditional) marginal distributions 
$\tilde{\epsilon}_s^t \given \B{Y} = \B{y} \sim \text{Uniform}(\{y_s^1, \dots, y_s^m\})$. It is worth noting that the variables
in $\tilde{\boldsymbol{\epsilon}}$ do not correspond to classical independent error terms in a regression model: 
they do not have mean zero ($\B{Y}$ has positive support), they are not mutually independent (they inherit all time-invariant spatial dependence in $\B{Y}$), 
and they are not independent of the other variables in the regression equation (roughly speaking, the dependence between
$\tilde{\boldsymbol{\epsilon}}$ and $\B{X}$ amounts to the dependence between $\B{Y}$ and $\B{X}$ that is induced via all
time-invariant confounders). Consequently, semi-synthetic datasets $(\B{X}, \tilde{\B{Y}})$ generated from the above model 
share the same time-invariant confounding as the original data. The true causal effect 
of $\B{X}$ on $\tilde{\B{Y}}$ is $T = f_{\text{AVE}(X \to Y)}(1) - f_{\text{AVE}(X \to Y)}(0) = \beta$.

\subsection{Power analysis (semi-real data)} \label{sec:power}
\begin{figure}[t]
\centering
\includegraphics[width=.5\linewidth]{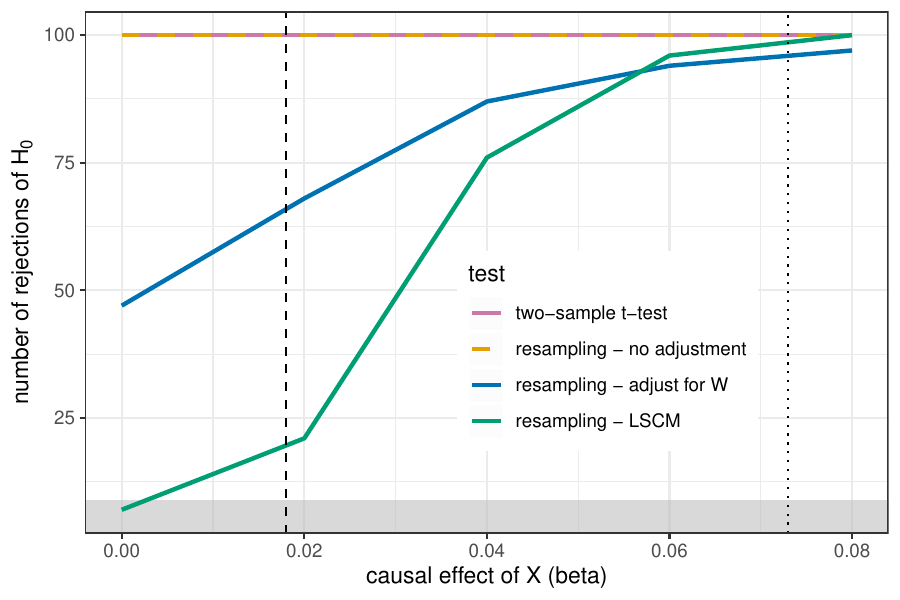}
\caption{Results of the power analysis described in Appendix~\ref{sec:power}. The colored curves show the empirial power
of the different methods. For increasing strength of the causal effect $\beta$, all tests tend to reject $H_0$. 
The vertical lines correspond to the absolute values of the estimated causal effect on the real data using our estimator (dashed), 
and the naive estimator based on difference in sample averages (dotted). If, indeed, the true causal effect was as large as suggested
by the naive analysis, our method (green curve) would detect it with high certainty. 
While the alternative procedures obtain higher detection rates, our method is the only one which holds the correct test level $\alpha = 0.05$:
for $\beta = 0$, only the green curve takes a value that lies within the lower $95\%$-quantile range of a $\text{Bin}(100, \alpha)$-distribution
(grey area).} 
\label{fig:power}
\end{figure}

Using the data generating process from Section~\ref{sec:semireal}, we generate, for each $\beta$ in the set $\{0, 0.02, 0.04, 0.06, 0.08\}$, 100 independent responses $\tilde{\B{Y}}_{\beta}^{(1)}, \dots, \tilde{\B{Y}}_{\beta}^{(100)}$ from \eqref{eq:genY}. 
For each resulting data set $(\B{X}, \tilde{\B{Y}}_{\beta}^{(i)})$, we compute a $p$-value for the hypothesis $H_0$ of a vanishing causal effect using 
the resampling test introduced in Section~\ref{sec:testing}. For comparison, we also include results for the two alternative testing procedures described in Section~\ref{sec:conflict_comparison}, 
which assume that there is no confounder, or that the only confounder is the distance to the closest road ($\B{W}$), respectively (see also Figure~\ref{fig:tests},
where we compare these three methods on the original data). In addition, we use a standard two-sample $t$-test, %
which tests whether $\E[Y_s^t \given X_s^t = 1] = \E[Y_s^t \given X_s^t = 0]$ (regarded as a test of the causal hypothesis $H_0$, this test assumes that 
there are no confounders).
For each $\beta$ and each testing procedure, we compute the number of rejections of $H_0$ (out of $100$) at level $\alpha = 0.05$.
The results are shown in Figure~\ref{fig:power}. 
As $\beta$ increases, our testing procedure (green curve) tends to reject $H_0$ more often.
For simulated data sets where the true causal effect is of the same magnitude as the estimated causal effect on the original data 
($T=-0.018$, the absolute value of which is shown by a vertical dashed line), we reject in roughly $20 \%$ of the cases. If the true effect is as large 
as suggested by the difference in sample averages on the original data ($T=0.073$, dotted vertical line), our test almost always rejects $H_0$. 
While the alternative approaches obtain higher rejection rates, none of them is able to keep the desired test level $\alpha = 0.05$ (see Figure~\ref{fig:power}).

\subsection{Bias due to a reduced sample size (semi-real data)} \label{sec:bias}
The estimator \eqref{eq:fAVEhat01} is constructed from a reduced dataset. The used data exclusion 
criterion is not independent of the assumed hidden confounders %
(i.e., the distribution of the hidden variables is expected to differ between the reduced 
data and the original data), 
and therefore potentially results in a biased estimator. %
Under additional assumptions on the underlying LSCM, however,
\eqref{eq:fAVEhat01} may still be used to estimate $T$. 
Below, we first give a population version argument, and then verify this argument empirically using a simulation experiment.

Assume that there is no interaction between $X_s^t$ and $H_s^t$ in the causal mechanism for $Y_s^t$, 
i.e., the function $f$ in \eqref{eq:f} splits into $f_1(X_s^t, \epsilon_s^t) + f_2(H_s^t, \epsilon_s^t)$. 
Then, the conditional expectation of $Y_s^t \given (X_s^t, H_s^t)$ likewise splits additively 
into a function of $X_s^t$ and a function of $H_s^t$. 
Using Proposition~\ref{prop:cov_adj}, 
it follows that any two different models for the marginal distribution of the latent process $\B{H}$
induce average causal effects $f_{\text{AVE}(X \to Y)}$ that are equal up to an additive constant. 
In particular, every model for $\B{H}$ induces the same value for $T$. 
By regarding the reduced dataset as a realization from a modified LSCM, 
in which the distribution of $\B{H}$ has been altered, 
this argument justifies the use of \eqref{eq:fAVEhat01} as an estimator for $T$.\footnote{In practice, 
we use $\B{X}$ to exclude data points; the above argument must thus be regarded a heuristic.}

\begin{figure}[t]
\centering
\includegraphics[width=\linewidth]{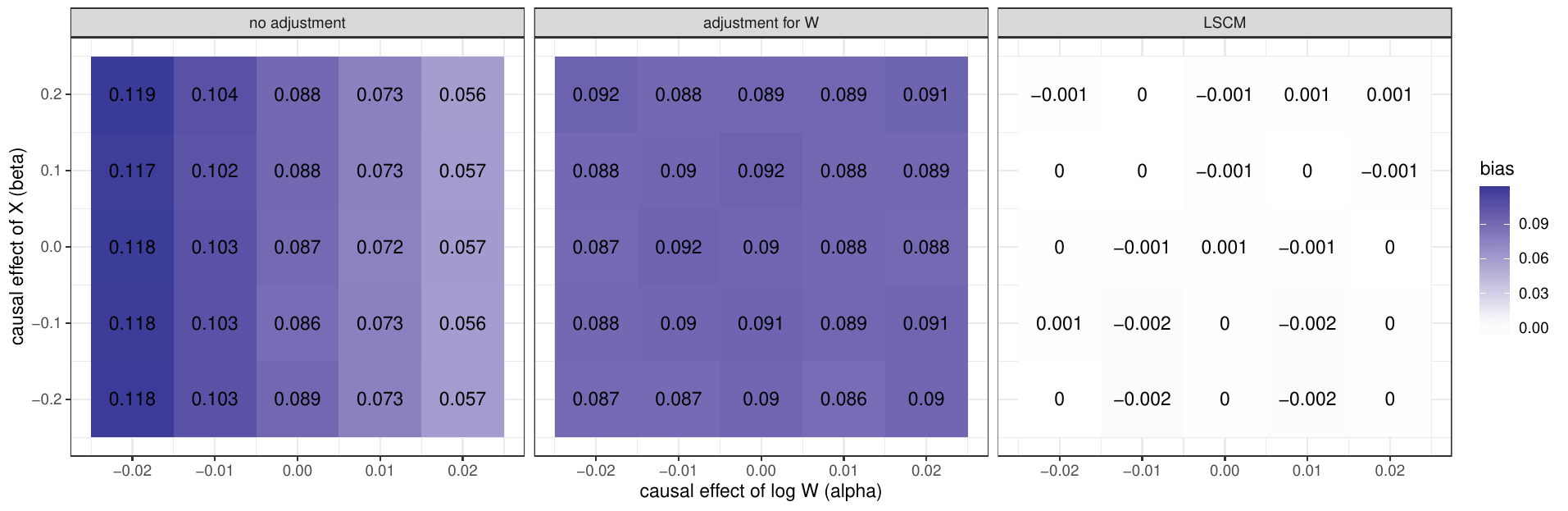}
\caption{Results of the bias analysis described in Appendix~\ref{sec:bias}. Along the first axis, we vary the causal effect of $ \log W_s^t$; along
the second axis that of $X_s^t$. For each parameter setting and each method, we compute an estimate $\hat \E [\hat T] - T$ of the bias, 
which is shown as numbers and color coding.
In these experiments, the causal effect of $X_s^t$ on $Y_s^t$ is 
confounded both by the observed variable $W_s^t$ and by unobserved time-invariant confounders (whose influence is transmitted via 
the error term $\tilde{\epsilon}_s^t$). The approach which does not adjust for confounding (left), and the approach which only adjusts for $W_s^t$
(middle), therefore lead to biased estimators. Our estimator (right) is approximately unbiased in all parameter settings. This property relies on 
the fact that, in the structural equations for $Y_s^t$, there is no interaction between $X_s^t$ and any of the confounders.}
\label{fig:bias}
\end{figure}

We now investigate the above statement empirically. To do so, 
we generate responses $\tilde{Y}_s^t$ from the model \eqref{eq:genY} in Section~\ref{sec:semireal}, but where we additionally include the additive term $\alpha \log W_s^t$,
where $\B{W}$ corresponds to the observed realization of the distance-to-road variable.%
\footnote{It is worth noting that for $\alpha \neq 0$, this model is not guaranteed to yield positive responses. 
We tried several alternative modeling approaches, too, 
which respect the distributional constraint $Y_s^t \geq 0$ (for example, we considered a zero-inflated log-normal regression model). 
Such approaches generally lead to poor model fits.}
Due to the highly skewed distribution of $\B{W}$, 
we model its influence on $\tilde{Y}_s^t$ on a log-scale. For datasets generated in this way, the causal effect of $X_s^t$ on $\tilde{Y}_s^t$
is confounded both by the dependence induced by $\tilde{\epsilon}_s^t$ (which includes all time-invariant confounding present in the original data), and by the dependence 
induced by the term $\alpha \log W_s^t$. 
For varying values of $\alpha$ and $\beta$, we simulate $100$ indendent responses 
$\tilde{\B{Y}}_{\alpha, \beta}^{(1)}, \dots, \tilde{\B{Y}}_{\alpha, \beta}^{(100)}$. On each dataset $(\B{X}, \tilde{\B{Y}}_{\alpha, \beta}^{(i)})$, we compute an estimate of 
the causal effect $T = \beta$ using our estimator \eqref{eq:fAVEhat01}. For comparison, we also include the two alternative estimators introduced in 
Section~\ref{sec:conflict_comparison}, which assume that there is no confounder, or that the only confounder is $\B{W}$, respectively (see also Figure~\ref{fig:estimator},
where we compare these three methods on the original data). For each choice of $\alpha, \beta$, and each method, we obtain an estimate
of the bias $\E[\hat{T}] - T$. The results are shown in Figure~\ref{fig:bias}. 
Both alternative approaches lead to biased estimators, since they either do not adjust for confounding, or only adjust for one of the confounders. 
Interestingly, the estimator shown in the left-most panel becomes less biased as the causal effect of $\log W_s^t$ increases. An explanation 
for this could be the following. We hypothesize that most time-invariant confounders introduce a positive correlation between $X_s^t$ and $Y_s^t$ (e.g., for 
a large distance to the closest road or for a small population density, both $X_s^t$ and $Y_s^t$ tend to take small values). As argued in Appendix~\ref{sec:semireal}, 
this dependence is preserved by our simulation procedure, and therefore confounds the causal effect of $X_s^t$ on $\tilde{Y}_s^t$ positively.
At the same time, the term $\alpha \log W_s^t$ introduces additional confounding correlation between $X_s^t$ and $\tilde{Y}_s^t$. For $\alpha < 0$, 
the additionally introduced correlation is positive 
(because we expect $W_s^t$ and $X_s^t$ to be negatively correlated),
and therefore adds to the overall confounding. For $\alpha > 0$, it is negative, and hence cancels out with some 
of the confounding introduced by $\tilde{\epsilon}_s^t$. After adjusting for $W_s^t$, the confounding reduces to what is induced by $\tilde{\epsilon}_s^t$
(indeed, the bias in the middle panel roughly coincides with the bias in the left panel for $\alpha = 0$). In all settings, our estimator (right panel) is approximately unbiased.

\section{Further results on resampling tests} \label{app:permutation}
\subsection{Temporal autocorrelation in the response variable}
A central assumption of the LSCM model class is that the error process 
of $\B{Y}$ is independent over time. This assumption says that all dependencies
between different temporal instances of $\B{Y}$ are induced via the covariates 
$\B{X}$ or the time-invariant confounders $\B{H}$. In practice, there may be 
other time-varying conditions influencing forest loss, thereby inducing a temporal 
dependence in $\B{Y}$ which cannot be explained by $(\B{X}, \B{H})$. In this case, 
the exchangeability property in Proposition~\ref{prop:exchange}, and therefore the 
level of our resampling test, is violated. To incorporate temporal autocorrelation in
the response variable, we adopt a block-permutation scheme: we divide the period 
2000--2018 into 6 blocks (2000--2002, 2003--2005, ..., 2016--2018), and 
perform a block-wise permutation of the data from $\B{Y}$. This procedure leaves the 
within-block dependence structure in $\B{Y}$ intact. The results align with our previous 
findings: $P = 0.892$ for the test of an instantaneous effect, and $P = 0.498$ for the 
test of a temporally lagged effect (when using the same test statistics as in Section~\ref{sec:conflict_results}).
\subsection{Spatial block-permutation scheme for Model~1}
In Section~\ref{sec:conflict_comparison}, we describe alternative permutation schemes 
to test the null hypotheses in Models~1~and~2. Strictly speaking, we require
additional exchangeability assumptions on $\B{Y}$ to ensure the validity of the 
corresponding resampling tests. Here, we investigate an alternative permutation
scheme for Model~1. To account for the spatial autocorrelation in $\B{Y}$, we adopt 
a spatial block-permutation: for every year 2000--2018, observations are grouped
into blocks of size $100 \si{\km} \times 100 \si{\km}$. To obtain resampled datasets, 
we then permute values of $\B{Y}$ in these blocks of data, thereby leaving the spatial 
dependence within each block intact. Observations which do not fall in any of the blocks
are permuted randomly. 
\begin{figure}
\centering
\includegraphics[width=.9\linewidth]{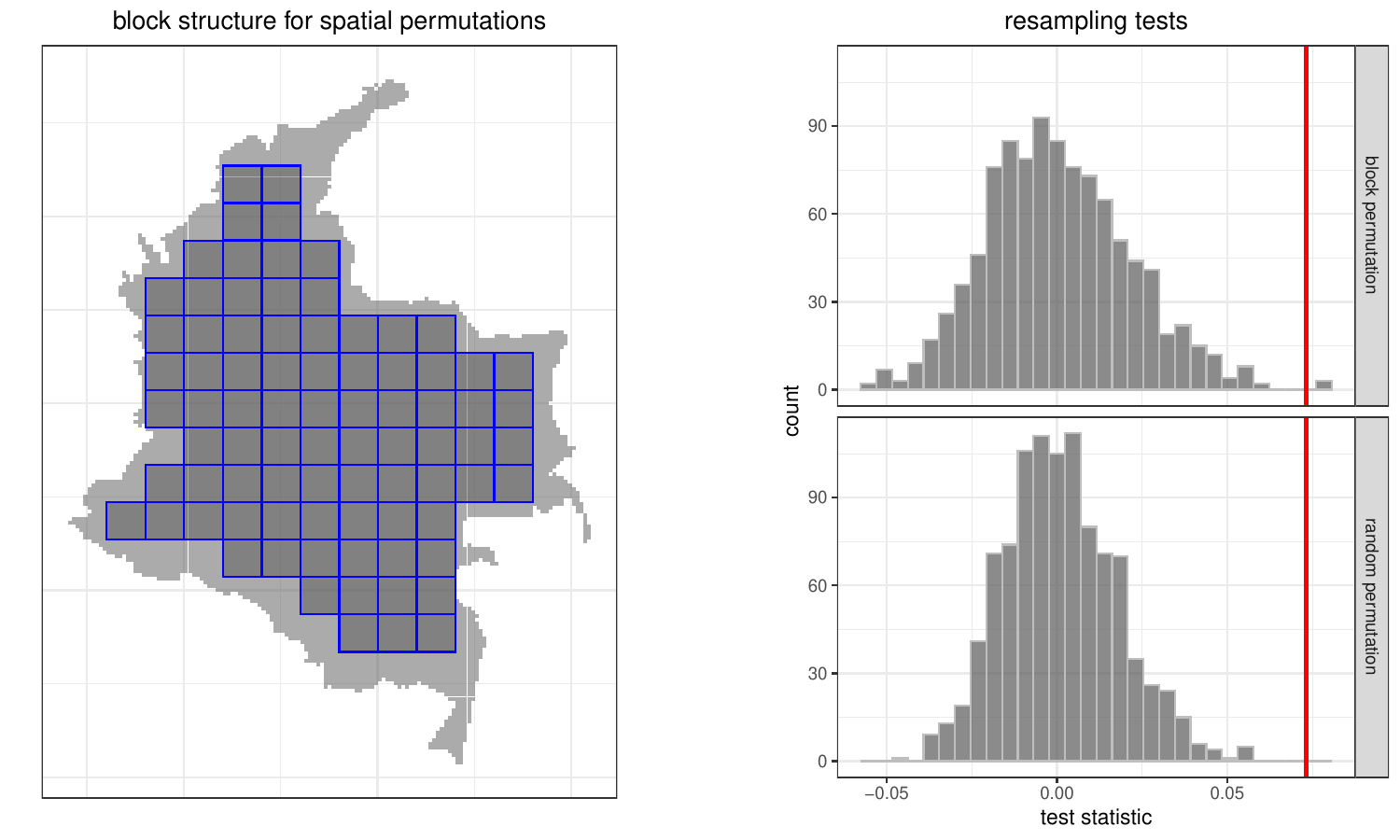}
\caption{
Block structure for the spatial permutation scheme (left) and results of resampling tests (right) for 
the null hypothesis in Model~1 from Section~\ref{sec:conflict_results}. The test statistic 
$\hat T = \hat{f}^{nm}_{\text{AVE}(X \to Y)}(1) - \hat{f}^{nm}_{\text{AVE}(X \to Y)}(0)$ is indicated by a red 
vertical bar.
The empirical distribution of the test statistic under this permutation scheme (top right) has a 
higher variance than under the permutation scheme used in Section~\ref{sec:conflict_results} (bottom right), 
resulting in a slightly larger $p$-value of $0.008$ compared with the $p$-value of $0.002$ for the 
original test. The significance of the test does not change.}
\label{fig:blockperm}
\end{figure}
As seen in Figure~\ref{fig:blockperm}, this procedure slightly 
increases the $p$-value, but does not affect the significance of the test.

\section{Proofs} \label{app:proofs}

\subsection{Proof of Proposition~\ref{prop:causal_int}}
By definition, intervening on $\B{X}$ leaves the conditional distribution $\B{Y} \given (\B{X}, \B{H})$ unchanged. Under $\P_x$, the property \eqref{eq:f} therefore still holds for the same error process $\boldsymbol{\epsilon}$. Since also the marginal distribution of $\B{H}$ is unaffected by the intervention, we have that 
\begin{align*}
\E_{\P_x} [Y_s^t] 	&= \E_{\P_x} [f(X_s^t, H^t_s, \epsilon_s^t)] 
							= \E_{\P_x} [f(x, H^t_s, \epsilon_s^t)] \\
							&= \E [f(x, H^t_s, \epsilon_s^t)]
							= \E [f(x, H^1_0, \epsilon_0^1)]
							= f_{\text{AVE}(X \to Y)}(x),
\end{align*}
as desired. $\hfill \square$

\subsection{Proof of Theorem~\ref{thm:consistency}}
Consider a fixed $x \in \mathcal{X}$. For every $n,m \in \N$ we have that 
\begin{align*}
\hat f^{nm}_{\text{AVE}(X \rightarrow Y)}(\B{X}_n^m, \B{Y}_n^m)(x) - f_{\text{AVE}(X \rightarrow Y)}(x)
&= \frac{1}{n} \sum_{i=1}^n \hat f^{m}_{Y \vert X} (\B{X}_{s_i}^m, \B{Y}_{s_i}^m)(x) - \E \left[ f_{Y \vert (X,H)} (x,H_0^1) \right] \\
&= \frac{1}{n} \sum_{i=1}^n \left( \hat f^{m}_{Y \vert X} (\B{X}_{s_i}^m, \B{Y}_{s_i}^m)(x) - f_{Y \vert (X,H)}(x,H^1_{s_i}) \right) \\
&\quad +\frac{1}{n} \sum_{i=1}^n f_{Y \vert (X,H)}(x,H^1_{s_i})- \E \left[ f_{Y \vert (X,H)}(x,H_0^1) \right] \\
&= r_1(\B{X}_n^m, \B{Y}_n^m, \B{H}_n^1) + r_2(\B{H}_n^1),
\end{align*}
where
\begin{align*}
r_1(\B{X}_n^m, \B{Y}_n^m, \B{H}_n^1) &:= \frac{1}{n} \sum_{i=1}^n \left( \hat f^{m}_{Y \vert X} (\B{X}_{s_i}^m, \B{Y}_{s_i}^m)(x) - f_{Y \vert (X,H)}(x,H^1_{s_i}) \right) \text{ and }\\
r_2(\B{H}_n^1) &:=  \frac{1}{n} \sum_{i=1}^n f_{Y \vert (X,H)}(x,H^1_{s_i})- \E \left[ f_{Y \vert (X,H)}(x,H_0^1) \right].
\end{align*}
It follows that for any $\delta > 0$,
\begin{align*}
\P \left( \left \vert \hat f^{n m}_{\text{AVE}(X \rightarrow Y)}(\B{X}_n^m, \B{Y}_n^m)(x) -  f^0_{\text{AVE}(X \rightarrow Y)}(x) \right \vert > \delta \right) 
&\leq \P \left( \vert r_1(\B{X}_n^m, \B{Y}_n^m, \B{H}_n^1) \vert > \delta / 2 \right) \\ 
&\quad + \P \left( \vert r_2(\B{H}_n^1) \vert > \delta / 2 \right).
\end{align*}
Let now $\alpha >0$ be arbitrary. By Assumption~\ref{ass:LLN}, there exists 
$N \in \N$ such that for all $n \geq N$, $\P(\card{r_2(\B{H}_n^1)} \geq \delta / 2) \leq \alpha / 2$. 
By Assumption~\ref{ass:fhat}, we can for any such $n \geq N$ find $M_n \in \N$, such that for all 
$i = 1, \dots, n$ and all $m \geq M_n$ it holds that $\P(\card{\hat f^{m}_{Y \vert X} (\B{X}_{s_i}^m, \B{Y}_{s_i}^m)(x) - f_{Y \vert (X,H)}(x,H^1_{s_i})} > \delta / 2) \leq \alpha / (2n)$. 
For all $m \geq M_n$ we then have 
\begin{align*}
\P \left( \card{r_1(\B{X}_n^m, \B{Y}_n^m, \B{H}_n^1)} > \delta / 2 \right) 
&\leq \P \left(   \frac{1}{n} \sum_{i=1}^n \left \vert \hat f^{m}_{Y \vert X} (\B{X}_{s_i}^m, \B{Y}_{s_i}^m)(x) - f_{Y \vert (X,H)}(x,H^1_{s_i}) \right \vert > \delta / 2 \right) \\
&\leq \P \left(   \bigcup_{i=1}^n \left \lbrace \left \vert \hat f^{m}_{Y \vert X} (\B{X}_{s_i}^m, \B{Y}_{s_i}^m)(x) - f_{Y \vert (X,H)}(x,H^1_{s_i}) \right \vert > \delta / 2 \right \rbrace \right) \\
&\leq \sum_{i=1}^n \P \left( \left \vert \hat f^{m}_{Y \vert X} (\B{X}_{s_i}^m, \B{Y}_{s_i}^m)(x) - f_{Y \vert (X,H)}(x,H^1_{s_i}) \right \vert > \delta / 2 \right) \\
&\leq \sum_{i=1}^n \alpha / (2n) = \alpha/2,
\end{align*}
and the result follows. $\hfill \square$

\subsection{Proof of Proposition~\ref{prop:LLN}}

By construction, $(H^1_{s_n})_{n \in \N}$ can be decomposed into $m$ subsequences 
$(H^1_{s_{(n-1)m + j}})_{n \in \N}$, $j \in \{1, \dots, m\}$, each of which corresponds to 
an equally spaced sampling of $\B{H}^1$ along the first spatial axis. 
We fist prove that each of these subsequences satisfies Assumption~\ref{ass:LLN}, 
and then conclude that the same must hold for the original sequence $(H^1_{s_n})_{n \in \N}$.
Let $j \in \{1, \dots, m\}$ and let $\varphi : \R^\ell \to \R$ be a measurable function with $\E[\card{\varphi(H_0^1)}] < \infty$. 
For notational simplicity, let for each $n \in \N$, $Z_n := H^1_{s_{(n-1)m + j}}$. 
The idea is to apply an ergodic theorem for real-valued stationary and ergodic time series 
\citep[e.g.,][Corollary~2.3.13]{sokol2013advanced} to the sequence $(\varphi(Z_n))_{n \in \N}$. 
By stationarity of the process $\B{H}^1$, and by choice of the sampling scheme, 
$(\varphi(Z_n))_{n \in \N}$ is indeed stationary. We need to show that $(\varphi(Z_n))_{n \in \N}$ is also ergodic. 
Using \cite[Lemma~2.3.15]{sokol2013advanced}, this follows by proving the following mixing condition: 
for all $p,q \geq 1$ and all $A_1, \dots, A_p \in \mathcal{B}(\R)$ and $B_1, \dots, B_q \in \mathcal{B}(\R)$, 
it holds that 
\begin{align} \label{eq:mixing}
\begin{split}
&\P(\varphi(Z_1) \in A_1, \dots, \varphi(Z_p) \in A_p, \varphi(Z_{n+1}) \in B_1, \dots, \varphi(Z_{n+q}) \in B_q) \\
\to \; &\P(\varphi(Z_1) \in A_1, \dots, \varphi(Z_p) \in A_p) \cdot \P(\varphi(Z_{1}) \in B_1, \dots, \varphi(Z_{q}) \in B_q) \quad \text{ as } n \to \infty.
\end{split}
\end{align}
Since the finite-dimensional distributions of $(Z_n)_{n \in \N}$ are Gaussian, this condition is easily verified. 
Let $p,q \geq 1$, and let $\P_1 = \mathcal{N}(\mu_1, \Sigma_1)$ and $\P_2 = \mathcal{N}(\mu_2, \Sigma_2)$ 
be the distributions of $(Z_1, \dots, Z_p)$ and $(Z_1, \dots, Z_q)$, respectively. 
Property \eqref{eq:mixing} follows if we can show that $(Z_1, \dots, Z_m, Z_{n+1}, \dots, Z_{n+p})$ 
converges to $\P_1 \otimes \P_2  = \mathcal{N}((\mu_1, \mu_2), \text{diag}(\Sigma_1, \Sigma_2))$
in distribution as $n \to \infty$. Convergence of the mean vector is trivial, and convergence of the 
covariance matrix follows by the assumption on $C$
and our choice of spatial sampling (the distance between the respective locations at which $(Z_1, \dots, Z_m)$ 
and $(Z_{n+1}, \dots, Z_{n+p})$ are observed tends to infinity as $n$ increases). To prove that the limit 
distribution is indeed Gaussian, one can then consider characteristic functions and apply a combination of 
Levy's Continuity Theorem \citep[e.g.,][Theorem~18.1]{williams1991probability} and the Cramér-Wold Theorem \citep{cramer1936some}. This proves that $\frac{1}{n} \sum_{i=1}^n \varphi(Z_i) \to \E[\varphi(Z_1)]$ 
in probability as $n \to \infty$, i.e., the subsequence $(H^1_{s_{(n-1)m + j}})_{n \in \N}$ satisfies Assumption~\ref{ass:LLN}. 
Since $j$ was arbitrary, this holds true for all $j \in \{1, \dots, m\}$. It remains to prove that also the original
sequence $(H^1_{s_n})_{n \in \N}$ satisfies Assumption~\ref{ass:LLN}.

Let an integrable function $\varphi : \R^\ell \to \R$ 
be given, and assume first that $\E[\varphi(H_0^1)] = 0$. 
For every $j \in \{1, \dots, m\}$ and $i \in \N$, define $S_i^j := \sum_{k=1}^i \varphi(H^1_{s_{(k-1)m+j}})$. 
By the first part of the proof, we have that for all $j$, $\tfrac{1}{i} S_i^j \to 0$ in probability as $i \to \infty$.
We want to show that also $\tfrac{1}{n} \sum_{k=1}^n \varphi(H^1_{s_k}) \to 0$ in probability as $n \to \infty$.
Let $\delta, \alpha > 0$ and choose $I \in \N$ such that for all $j \in \{1, \dots, m\}$ and $i \geq I$, $\P(\frac{1}{i} \card{S_i^j} > \delta / m) \leq \alpha / m$. 
Define $N := m I+1$ and pick an arbitrary $n \geq N$. 
We can then write $n = i m + j$ for some $i \geq I$ and $j \in \{1, \dots, m\}$. 
With $J_1 := \{1, \dots, j\}$ 
and $J_2 = \{1, \dots, m\} \setminus J_1$, we then have
\begin{align*}
\P ( \frac{1}{n} \vert \sum_{k=1}^n \varphi(H^1_{s_k}) \vert  > \delta ) &= \P(\frac{1}{n} \vert \sum_{{j^\prime} \in J_1} S_{i+1}^{j^\prime}+ \sum_{{j^\prime}  \in J_2} S_{i}^{{j^\prime}} \vert > \delta ) \\ &\leq \P(\sum_{{j^\prime}  \in J_1} \frac{1}{i+1} \card{S_{i+1}^{j^\prime} }+ \sum_{{j^\prime}  \in J_2} \frac{1}{i} \card{S_{i}^{j^\prime} } > \delta ) \\ &\leq \sum_{{j^\prime}  \in J_1} \P(\frac{1}{i+1} \card{S_{i+1}^{j^\prime} } > \delta / m) + \sum_{{j^\prime}  \in J_2} \P(\frac{1}{i} \card{S_{i}^{j^\prime} } > \delta / m) \leq \alpha,
\end{align*}
which completes the proof in the case where $\E[\varphi(H_0^1)] = 0$. 
The general case follows by applying the above result to the function $\tilde \varphi = \varphi - \E[\varphi(H_0^1)]$. 
\hfill $\square$

\subsection{Proof of Proposition~\ref{prop:fhat}}

Let $s \in \R^2$ and $\B{h} \in \mathcal{H}$. 
With $\gamma_s := f_1(h_s^1)$, it follows from (L1) that for all $x$ and $t$, we have $\E[Y_s^t \given X_s^t = x, H_s^t = h_s^t] = \varphi(x)^\top \gamma_s$. 
It therefore suffices to prove that $\hat{\gamma}_s^m \to \gamma_s$ in probability under $\P_\B{h}$. 
For the ease of notation, we omit all sub- and superscripts from $\B{Y}_s^m, \B{\Phi}_s^m$ and $\boldsymbol{\xi}_s^m$ in the below calculations. 
Let $c > 0$ be such that (L3) holds true and let $\delta > 0$ be arbitrary. For every $m \in \N$, we then have
\begin{align*}
\P_\B{h}(\norm{\gamma_s - \hat \gamma_s^m}_2 > \delta) 	&= \P_\B{h}(\norm{\gamma_s -  (\B{\Phi}^\top \B{\Phi})^{-1} \B{\Phi}^\top \B{Y}}_2 > \delta) \\
																														&= \P_\B{h}(\norm{\gamma_s -  (\B{\Phi}^\top \B{\Phi})^{-1} \B{\Phi}^\top (\B{\Phi} \gamma_s + \boldsymbol{\xi})}_2  > \delta) \\
																														&= \P_\B{h}(\norm{(\B{\Phi}^\top \B{\Phi})^{-1} \B{\Phi}^\top \boldsymbol{\xi}}_2 > \delta) \\
																														&\leq \P_\B{h}(\norm{( \tfrac{1}{m}\B{\Phi}^\top \B{\Phi})^{-1}}_2 \norm{ \tfrac{1}{m} \B{\Phi}^\top \boldsymbol{\xi} }_2 > \delta \given) \\
																														&= \P_\B{h}((\lambda_\text{min}( \tfrac{1}{m}\B{\Phi}^\top \B{\Phi}))^{-1} \norm{ \tfrac{1}{m} \B{\Phi}^\top \boldsymbol{\xi} }_2 > \delta) \\
																														&= \P_\B{h}((\lambda_\text{min}( \tfrac{1}{m}\B{\Phi}^\top \B{\Phi}))^{-1} \norm{ \tfrac{1}{m} \B{\Phi}^\top \boldsymbol{\xi} }_2 > \delta \text{ and } \lambda_{\text{min}} \left( \tfrac{1}{m} \B{\Phi}^\top \B{\Phi} \right) > c  ) \\
																														&\quad + \P_\B{h}((\lambda_\text{min}( \tfrac{1}{m}\B{\Phi}^\top \B{\Phi}))^{-1} \norm{ \tfrac{1}{m} \B{\Phi}^\top \boldsymbol{\xi} }_2 > \delta \text{ and } \lambda_{\text{min}} \left( \tfrac{1}{m} \B{\Phi}^\top \B{\Phi} \right) \leq c  ) \\
																														&\leq \P_\B{h}(\norm{ \tfrac{1}{m} \B{\Phi}^\top \boldsymbol{\xi} }_2 > c \delta) 
																														+ \P_\B{h}(\lambda_{\text{min}} \left( \tfrac{1}{m} \B{\Phi}^\top \B{\Phi} \right) \leq c), \\
\end{align*}
which tends to zero as $m \to \infty$ by (L2) and (L3). \hfill $\square$

\subsection{Proof of Proposition~\ref{prop:exchange}}
Recall our definition of $\mathcal{H} \subset \mathcal{Z}_\ell$ as the set of functions $\B{h} : \R^2 \times \N \to \R^\ell$ that are constant in the
time-argument. Since $\B{H}$ is time-invariant, we have that $\P(\B{H} \in \mathcal{H}) = 1$. It therefore suffices to prove that for all $\B{h} \in \mathcal{H}$,
$(\B{X}_n^m, \sigma(\B{Y}_n^m)) \stackrel{d}{=} (\B{X}_n^m, \B{Y}_n^m)$ under $\P( \, \cdot \given \B{H} = \B{h})$. 
Assume that $H_0$ holds true, and let $\sigma$ be a permutation of $\{1, \dots, m\}$.
Then, there exists a function $\tilde{f} : \R^{\ell+1} \to \R$ and an error process 
$\boldsymbol{\epsilon} \indep (\B{X}, \B{H})$ such that for all $(s,t) \in \R^2 \times \N$, $Y_s^t = \tilde{f}(H_s, \epsilon_s^t)$. 
It follows that the conditional distribution of $\B{Y} \given (\B{X}, \B{H})$ does not depend on $\B{X}$, 
and hence that $\B{X}$ and $\B{Y}$ are conditionally independent given $\B{H}$. 
Furthermore, since $\boldsymbol{\epsilon}^1, \boldsymbol{\epsilon}^2, \dots$ are i.i.d., we have that 
for all $\B{h} \in \mathcal{H}$, 
$\B{Y}^1, \dots, \B{Y}^m$ are i.i.d.\ under $\P( \, \cdot \given \B{H} = \B{h})$. 
For all $\B{h} \in \mathcal{H}$, it therefore holds that $(\B{X}_n^m, \sigma(\B{Y}_n^m)) \stackrel{d}{=} (\B{X}_n^m, \B{Y}_n^m)$
under $\P( \, \cdot \given \B{H} = \B{h})$, and the result follows. $\hfill \square$

\bibliography{../ref}

\end{document}